\documentclass[a4paper,12pt]{article}

\usepackage{amsmath}
\usepackage{amssymb}
\usepackage{amsfonts}
\usepackage{amsthm}

\usepackage{palatino}
\usepackage{graphics}
\usepackage{graphicx}
\usepackage{lscape}

\usepackage{appendix}
\usepackage[bookmarks]{hyperref}
\usepackage{caption}
\usepackage{subcaption}
\usepackage{makecell}
\usepackage{booktabs}
\usepackage{longtable}

\usepackage{natbib}

\usepackage{todonotes}

\usepackage[top=2.0cm, bottom=2.0cm, outer=2.0cm, inner=2.0cm, heightrounded]{geometry}

\usepackage{adjustbox}

\setuptodonotes{inline,size=small,fancyline, color=blue!30}

\usepackage{setspace}

\newcolumntype{H}{>{\setbox0=\hbox\bgroup}c<{\egroup}@{}}

\begin{document}

\title{(In)stability in the Dynamics of the Cross-Country Distribution of Income Per Capita}
\author{Davide Fiaschi\thanks{Dipartimento di Economia e Management, University of Pisa, Via Ridolfi 10, 56124 Pisa, Italy} \and Paul Johnson\thanks{Department of Economics, Vassar College, 124 Raymond Ave, Poughkeepsie, NY 12604, USA}}
\date{June 6, 2025}

\maketitle

\pagenumbering{gobble}

\begin{abstract}

	\singlespacing{Using a panel of 102 countries from PWT 10.0 covering 1970-2019, we examine the veracity of the assumption that a time-homogeneous, first-order process describes the evolution of the cross-country distribution of per capita output, an assumption often made in studies of the convergence hypothesis employing the distribution dynamics approach pioneered by \citet{quah1993empirical}. To test homogeneity, we compare transition kernels estimated for different time periods and, for those periods exhibiting evidence of homogeneity, we test the first-order assumption using an implication of such a process’s Chapman-Kolmogorov equations. Both tests require measurement of the distance between probability distributions which we do with several different metrics, employing bootstrap methods to assess the statistical significance of the observed distances. We find that the process was time-homogeneous and first-order in the 1970-1995 period during which the distribution dynamics imply a bimodal long-run distribution, consistent with convergence clubs. Following the apparent break in the process in the late 1990s, the 2000-2010 distribution dynamics imply a unimodal long-run distribution suggestive of a single convergence club, consistent with recent claims of short-term $\beta$-convergence from the late 1990s and beyond made by \citet{patel2021new} and \citet{kremer2022converging}. After 2010, there is some evidence of a return to non-convergent dynamics similar to those of the 1970-1995 period.}

\end{abstract}

\bigskip
\bigskip


\bigskip
\bigskip
\bigskip

{\small \noindent \textbf{Aknowledgements: }
We thank Michele Battisti, Fulvio Corsi, Margherita Gerolimetto, Daniel Henderson, Andrea Mario Lavezzi, Stefano Magrini, Angela Parenti, and Cristiano Ricci for their comments on an earlier version.
Johnson thanks the Dipartimento di Economia e Management at the University of Pisa for their hospitality during several visits.}

\newpage
\tableofcontents
\newpage

\clearpage

\pagenumbering{arabic}
\setcounter{page}{1}

\section{Introduction}

Early studies of the convergence hypothesis in the modern empirical growth literature typically define convergence as catching-up behaviour based on the relationship between initial per capita income and its subsequent growth (so-called $\beta$-convergence) or as a reduction in the cross-country dispersion of per capita income over time (so-called $\sigma$-convergence).\footnote{\citet{durlauf2009rise} discusses earlier interest in the convergence or divergence of nations. \citet{durlauf1999}, \citet{islam2003}, \citet{durlauf2005}, \citet{johnson2020remains}, and \citet{johnsonEtAl2024} provide surveys of the convergence literature.}
In a highly influential series of papers, \citet{quah1993empirical,quah1996convergence,quah1996empirics,quah1996twin, quah1997empirics,quah1999} criticizes these approaches to testing the convergence hypothesis as being unsuitable for the study of key issues such as mobility, stratification, and polarization in the cross-country distribution of per capita output. As an alternative, he proposes ``distribution dynamics'': studying the evolution of the entire cross-country distribution of per capita output rather than just a single parameter implied by that evolution.

\citet{quah1993empirical} implements the distribution dynamics approach by discretizing the state space of the cross-country distribution of per capita GDP and using a Markov chain to model the evolution of the distribution and to examine its long-run implications. However, following \citet{quah1996convergence,quah1997empirics}, most recent applications of distribution dynamics employ almost exclusively continuous state space methods.\footnote{The motivation for a continuous-state-space approach is that the discretization of the state space of a continuous-state-space Markov process can damage the Markov property. See \citet{billingsley1961}, \citet{kemeny2012denumerable} and \citet{guihenneuc1998discretization} regarding this issue which is discussed in the economic growth context by \cite{quah1996convergence,quah1997empirics,quah2001searching}, \citet{magrini1999}, \citet{reichlin1999}, and \citet{bulli2001}. Applications are surveyed in \citet{durlauf2005}, \citet{johnson2020remains}, and \citet{johnsonEtAl2024}.}
These studies provide evidence of considerable persistence and a strong long-run tendency to multimodality in the cross-country distribution of per capita output, consistent with the club convergence hypothesis. Their analysis typically assumes that the process describing the evolution of the cross-country distribution of per capita output is: (i) homogenous; and, (ii) first-order. Consequently, at least some of our understanding of the dynamics of the cross-country distribution of per capita output rests on these two assumptions. However, despite their importance, little attention has been given to assessing their veracity in the economic growth context.\footnote{The first-order assumption also arises in other contexts in the growth literature. For example, the often-used log-linear approximation to the neoclassical growth assumes a first-order process so conclusions based on the approximation could be erroneous if the true process is of second or higher order which would admit potentially richer dynamics such as different speeds of convergence, possibly non-monotonic convergence, and possible overshooting of the long-run distribution. \citet[p591-2]{durlauf2005} discuss the literature examining the accuracy of the first-order linear approximation to the law of motion in the one sector neoclassical growth model.}

Apart from the statistical concern that the modelling of data generated by a process that does not satisfy both the homogeneity and first-order assumptions constitutes a specification error that could invalidate any inferences drawn from the analysis, there is reason for interest in the general stability of the evolution of per capita GDP. Recently, \citet{patel2021new} and \citet{kremer2022converging} have presented evidence of short-term $\beta$-convergence in cross-country data from the late 1990s and beyond.\footnote{ \citet{johnsonEtAl2024} criticize the short-term $\beta$-convergence findings on the grounds that (i) $\beta$-convergence tests have low power against the important club-convergence alternative which has strong support in the literature; and, (ii) a data span of 10 years is too short to study convergence which is a long-run concept.} For example, \citet{kremer2022converging} present evidence of a statistically significant negative correlation between initial per capita GDP and its growth over the subsequent 10-year period when the initial period is 2000. This result is at variance with the standard finding that such correlation is generally either positive or not significantly different from zero when earlier initial periods or longer time spans are considered \citep{johnson2020remains}. That difference could reflect a change (i.e. an inhomogeneity) in the process describing the evolution of the cross-country distribution of per capita GDP and so investigation of the homogeneity hypothesis, and any analysis of the nature of any inhomogeneity found, could throw some light on the forces behind, and implications of, the short-term $\beta$-convergence claims.

To the best of our knowledge there are no tests, formal or informal, of the homogeneous, first-order specification of the evolution of the cross-country distribution of per capita GDP in the continuous-state space case. The objective of this paper is to provide and apply such tests. The tests are: (i) a test of homogeneity which is a continuous state space analogue of the \citet{bickenbach2003} tests of homogeneity in that it compares the estimated one-period transition operators (the analogue of the transition matrix) for different time periods; and, (ii) a test of the first-orderedness of the process based on the Chapman-Kolmogorov equations for the process, which compares the two-period iteration of the estimated one-period transition operator with the estimated two-period transition operator. In both cases the tests involve comparing the distance between two estimated distributions with the sampling distribution of that distance which we do using bootstrap methods. As there is no unique way to measure the distance between two distributions we use Monte Carlo methods to investigate the effects of the metric used on the size and power of our tests.

One important innovation in this paper is that we acknowledge that the length of the transition period is unknown. While we have annual data, there is no reason to assume that one year is the transition length of the underlying growth process which could be 5 years, 10 years, 15 years, and so on. Our approach to this issue is to perform the tests described above to data for a variety of transition lengths, looking for that, if any, for which the process satisfies the homogeneous, first-order specification.

Before describing our testing framework in Section 3, the next section discusses our dataset and uses that data to provide some preliminary evidence of the stability and in stability in the cross-country distribution of per capita GDP.

\section{Data and motivating evidence}
\label{sec:datamotivation}

In this section, we discuss our data and take a preliminary look at it in order to further motivate the testing procedure that we then outline and implement. Our data are a balanced sample of 102 countries over the 1970-2019 period from PWT 10.0 \citep{feenstra2015}. Using the PWT mnemonics, we measure output per capita as RGDPO/EMP. To construct the balanced sample we exclude all countries without data on RGDPO and/or EMP for the entire period as well as the middle-eastern oil producing countries and (supposed) tax heavens. Starting our data in 1970 gives a good balance between the cross-section and time dimensions of the dataset. The data are normalized by dividing by the average value of RGDPO/EMP across the 102 countries in each year.\footnote{The PWT codes of the 102 countries in the sample are  AGO, ALB, ARG, AUS, AUT, BEL, BFA, BGD, BGR, BOL, BRA, BRB, BWA, CAN, CHE, CHL, CHN, CIV, CMR, COD, COG, COL, CRI, CYP, DEU, DNK, DOM, DZA, ECU, EGY, ESP, ETH, FIN, FRA, GAB, GBR, GHA, GRC, GTM, HKG, HND, HTI, HUN, IDN, IND, ISL, ISR, ITA, JAM, JOR, JPN, KEN, KHM, KOR, LBN, LCA, LKA, MAR, MDG, MEX, MLI, MLT, MMR, MOZ, MUS, MWI, MYS, NAM, NER, NGA, NLD, NOR, NZL, PAK, PAN, PER, PHL, POL, PRT, PRY, ROU, RWA, SDN, SEN, SGP, SWE, SYR, TCD, THA, TTO, TUN, TUR, TWN, TZA, UGA, URY, USA, VEN, VNM, ZAF, ZMB, and ZWE. Dataset and codes are available at \url{https://people.unipi.it/davide_fiaschi/research/}.}

\begin{figure}[htbp]
		\centering
		\centering
		\includegraphics[width=0.6\textwidth]{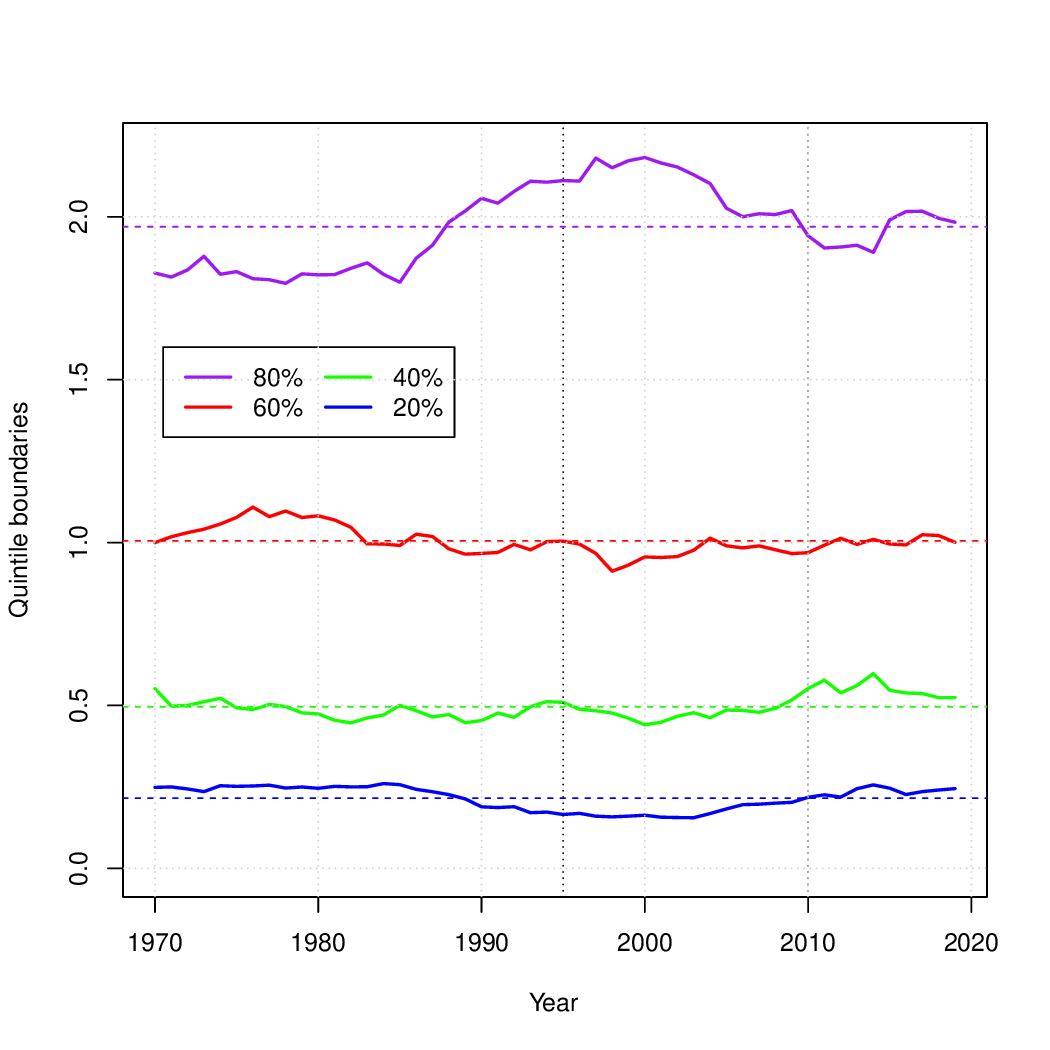}
		\caption{Quintile boundaries for each year in the sample. Indicated is the fraction of the sample falling below each line. The dashed lines indicate the sample median value of the quintile boundary of the same colour. Authors' calculations using data from \citet{feenstra2015}.}
		\label{fig:normalizedQuintileBoundaries}
\end{figure}

Figure \ref{fig:normalizedQuintileBoundaries} shows the evolution of the quintile boundaries of the data over the sample period.

\begin{figure}[htbp]
		\centering
		\includegraphics[width=0.6\textwidth]{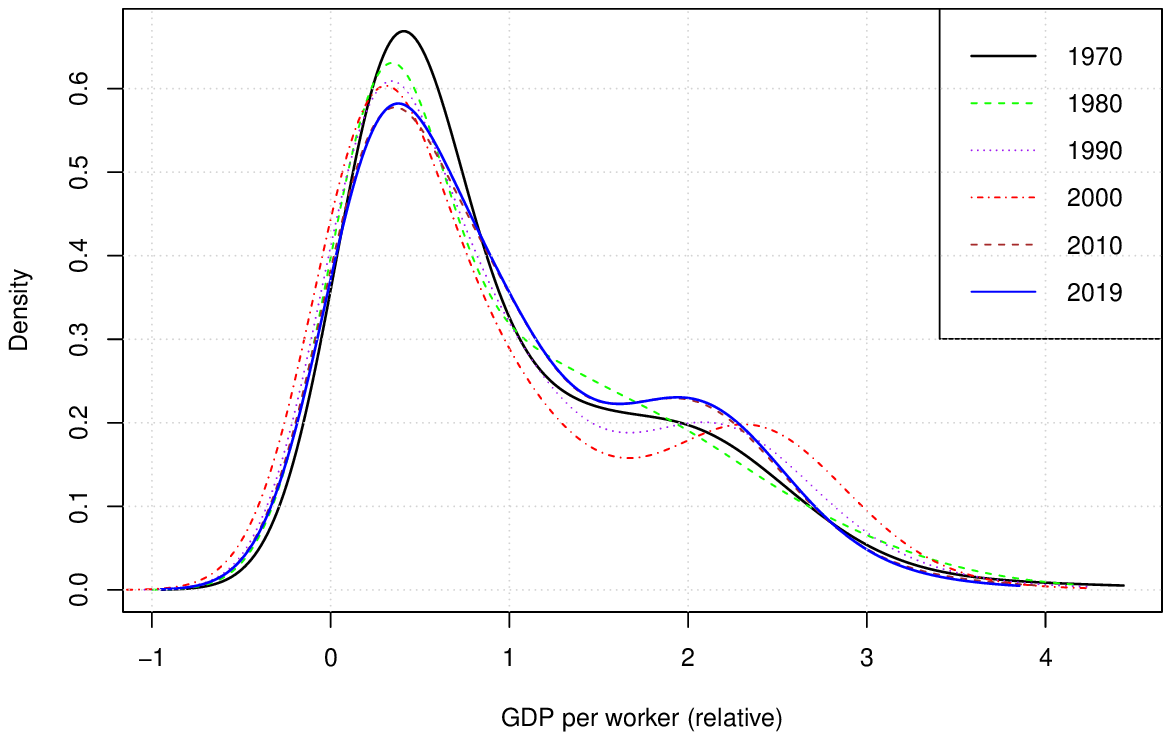}
		\caption{The estimated cross-sectional densities of (relative) GDP per worker. The densities are estimated from the sample of 102 countries described in the text, for the years 1970, 1980, 1990, 2000, 2010 and 2019, using a adaptive kernel density estimator with a Gaussian kernel and the optimal normal bandwidth
		\citep[p101]{silverman86}. Authors' calculations using data from \citet{feenstra2015}.}
		\label{fig:estimatedCrossSectionalDistributionsDifferentYears}
\end{figure}

The most striking features of Figure \ref{fig:normalizedQuintileBoundaries} are the approximate constancy, relative to the respective medians indicated by the dashed lines, of the boundaries between the bottom three quintiles (the 20\% and 40\% lines) and, to a slightly lesser extent, that between the third and fourth quintile (the 60\% line), as well as the increase in the gap between the top quintile and the rest of the sample (the 80\% line) from the mid-1980s through 2000 and its subsequent fall during the 2000s to a level similar to that in the earlier part of the sample period. These features, especially the relative volatility in the 80\% line, hint at stability in the evolution of the cross-country distribution of per capita GDP early in our sample, followed by instability, especially among the higher income countries, which then moderates towards the end of the sample.

Figure \ref{fig:estimatedCrossSectionalDistributionsDifferentYears} shows the estimated cross-sectional densities for the data set in 1970, 1980, 1990, 2000, 2010 and 2019 computed using a adaptive kernel density estimator with a Gaussian kernel and the optimal normal bandwidth \citep[p101]{silverman86}.\footnote{Here and in our tests present below, we do not make any adjustments to the estimation procedure to account for the boundary of the support of the distribution of the cross-country distribution of per capita GDP at zero. Doing so using the reflection method described in \citep[Section 2.10]{silverman86}, for example, constrains the derivative of the estimated density to be zero at zero, which may not be true. Rather, we rely on the adaptive kernel estimator, which uses a smaller bandwidth in places where the data is dense, and the relatively large number of countries with low to middle levels of per capital GDP, to minimize the amount of mass falling below zero in the estimated densities.} Immediately evident is the well-known bimodality of the cross-country distribution of per capita GDP, the "emerging twin peaks" discussed by \citet{quah1997empirics}. Also evident are shifts in the distributions consistent with the fluctuations in the quintile boundaries observed in Figure \ref{fig:normalizedQuintileBoundaries}. For example, the estimated densities for 1980, 1990, and 2000 are similar below the (relative) mean value of one. However, the estimated density for 1990 has more mass above twice the mean than does that for 1980, while that for 2000 has less mass near the antimode than that for 1990, and more mass above the right-hand mode which lies to the right of the estimated 1990 right-hand mode. That is, between 1980 and 2000, there is a rightward shift of the mass above the mean in the estimated densities, consistent with the rise in the gap between the 60\% and 80\% quintile boundaries evident in Figure 1. This shift is then reversed resulting in the the upper tails of the estimated densities for 2010 and 2019 being to the left of that for 2000. As with Figure \ref{fig:normalizedQuintileBoundaries}, the impression is of stability in the first part of the sample followed by instability, especially among the higher income countries, which then moderates towards the end of the sample.

The behaviour of the cross-country distribution of per capita GDP observed in Figures \ref{fig:normalizedQuintileBoundaries} and  \ref{fig:estimatedCrossSectionalDistributionsDifferentYears} is also relevant to the claims of short-term $\beta$-convergence in the 2000s reported by \citet{patel2021new} and \citet{kremer2022converging} in as much as the reduction in the gap between the top and bottom of the distribution seems rather more due to a downward movement (relative to the mean) of the countries in the top 20\% of the distribution than to an upward movement of those in the bottom 60\%. More generally, the prima facie evidence of instability is potentially inconsistent with the “monotonic” convergence to an equilibrium distribution implied by the neoclassical growth model, for example. Plausible explanations for the observed behaviour include more complex convergence behaviour, shocks that affect counties in different parts of the distribution to greater or lesser extents, and instability in the dynamic process governing the evolution of the distribution.

Our objective here is to investigate the stability of the process and, while the next section outlines our approach to doing so, it is worth looking ahead to our results which offer support for the hypothesis that the dynamics of the distribution of (relative) GDP per worker were governed by an homogeneous, first-order process in the 1970-1995 period. That process is characterized by a two-peaked distribution of (relative) GDP per worker in the short and in the long run. However, there is strong evidence of a large shift in the process in the mid- to late-1990s which ushers in a decade of dynamics broadly consistent with a single-peaked long-run distribution before an apparent return to dynamics similar to those in the first part of the sample.

The next section outlines the continuous state space distribution dynamics approach and gives the details of the specification tests that we propose, including the bootstrapping that we employ for inference.

\section{Continuous state space distribution dynamics and specification tests  \label{sec:introductionContinuosStateSpace}}

\citet{quah1996convergence,quah1997empirics} develops and advocates a continuous state space approach to ``distribution dynamics'', the modelling of the evolution of the cross-country distribution of per capita income. To outline this approach, we assume that the time-$t$ cross-country distribution of a variable $x$ can be described by the density function $f_t(x)$.\footnote{While the fundamental idea here is the same as that in \citet{quah1996convergence,quah1997empirics} we have simplified the exposition by assuming that the marginal and conditional income distributions have density functions. Quah's development of the approach avoids these assumptions and is far more general. See \citet{durlauf1999} for a through exposition of the more general approach.}

In general, this distribution will evolve over time so that the density prevailing at time $t+\tau$ for $\tau>0$ is $f_{t+\tau}(x)$. Under the assumption that the process describing the evolution of the distribution is time homogeneous and first-order, the relationship between the two densities can be written as
\begin{equation}
	f_{t+\tau}(z)=\int_\mathbb{R}g_\tau(z|x)f_t(x)dx,
	\label{eq:distributionDynamics}
\end{equation}
where $g_\tau(z|x)$ is the $\tau$-period-ahead density of $z$ conditional on $x$ i.e. for any fixed value of $x$, the function $g_\tau(z|x)$ is analogous to a row of a Markov transition matrix for a discrete state space process. So long as it exists, the ergodic (long-run) density, $f_\infty(z)$, implied by $g_\tau(z|x)$ can be found as the solution to $f_\infty(z)=\int_\mathbb{R}g_\tau(z|x)f_\infty(x)dx$.\footnote{\citet[p364]{judd1998} provides a method of solving this equation to compute the ergodic density.}

\citet{quah1996convergence,quah1997empirics} applies this approach to data on the cross-country distribution of per capita output from 1961 to 1988 with $\tau=15$ and finds strong evidence of persistence within the distribution as well as evidence of a long-run tendency towards bimodality or ``emerging twin-peaks'' in the distribution, contrary to the convergence hypothesis.\footnote{ Persistence is evidenced by the finding that, for all values of $x$, much of the probability mass in the estimated $g_\tau(z|x)$ is concentrated around values of $z$ with $z\approx x$. The tendency towards bimodality is evidenced by lower concentrations of mass around such values of $z$ for middling values of $x$ than for both low and high values of $x$ implying a tendency for countries to gravitate towards either high or low values of per capita output (see Figures 5.1 and 5.2 in \citealp{quah1997empirics}). \citet{johnson2005} computes the ergodic density implied by an estimated $g_\tau(z|x)$ and confirms the long-run tendency to bimodality in the cross-country distribution of per capita output.}

A large body of research has employed these methods including \citet{andres1995}, \citet{fiaschi2003distribution,lavezzi2003determinants, fiaschi2007nonlinear, fiaschi2007appropriate, fiaschi2020deep}, \citet{johnson2005}, \citet{fotopoulos2006nonparametric,fotopoulos2008european}, \citet{maasoumi2007growth}, \citet{fischer2008income}, \citet{bandyopadhyay2004twin, bandyopadhyay2011rich}, \citet{barseghyan2011cross}, in addition to the related research cited in \citet{durlauf2005}, \citet{johnson2020remains}, and \citet{johnsonEtAl2024}.
Following \citet{quah1996convergence,quah1997empirics}, this body of research typically assumes that the process describing the evolution of the cross-country distribution of per capita output is: (i) homogenous; and, (ii) first-order. Consequently, at least some of our understanding of the dynamics of the cross-country distribution of per capita output rests on these two assumptions. However, despite their importance, little attention has been given to assessing the veracity of these assumptions for a continuous state space process in the economic growth context.

In the discrete state space case there has been some study of the applicability of the first-order time homogeneous Markov chain. \citet{fingleton1997} tests the homogeneity assumption by modelling the transition probabilities as log-linear functions of the origin and destination states and the time period. Rejecting the hypothesis that the coefficient on the time variable is zero, he rejects homogeneity. \citet{bickenbach2003} test the homogeneity and first-orderedness assumptions with data on relative per capita incomes from the 48 contiguous US states over the period 1929-2000 by applying the tests of \citet{anderson1957}. They find some evidence of homogeneity for the 1950-1995 period and some evidence supporting the first-order hypothesis when biannual transitions are considered. Using data on cross-country distribution of per capita GDP, \citet{quah1993empirical} performs an informal test of the homogeneous, first-order specification in the discrete-state space case based on the Chapman-Kolmogorov equations for the process which imply that the $r$-period transition matrix for a first-order process equals the one-period transition matrix raised to the $r^{\text{th}}$ power. Quah thus compares a transition matrix estimated for the 23-year period 1962-1985 with that estimated for annual transitions over the same period raised to the 23$^{\text{rd}}$ power. He finds that the two matrices differ in that the latter understates the persistence in the cross-country income distribution implied by the former but is unable to say whether or not this difference is within the bounds of sampling error.\footnote{\citet{shorrocks1976income} reports the same phenomenon for transition matrices estimated from individual income data and offers some possible reasons. \citet{quah1993empirical} is careful to verify that the bimodality in the ergodic distributions implied by his estimated transition matrices is robust to a relaxation of the assumption of a first-order process to that of a second-order process.} Similarly, \citet{kremer2001searching} compare the square of an estimated 5-year transition matrix with an estimated 10-year transition matrix but do not formally test their equality.$\,$

\citet{fiaschiJohnson2023b} provide formal specification tests of the first-order Markov model of the evolution of the cross-country distribution of per capita income and find that, while there is some evidence of inhomogeneity for transition lengths of 5 and 10 years, cross-country data for the 1970-2019 period are consistent with the first-order Markov model provided that the transition period is taken to be 15 years or longer. Using a transition length of 4 years, \citet{imam2025} model cross-country transition dynamics using a Markov chain model with the crisis regimes in addition to the states defined by income levels. They find evidence of an inhomogeneity in the dynamics in the mid-1990s which they suggest may reflect a reduction in the number and severity of crises.

The tests that we propose have two parts. We first compare the estimated $g_\tau(z|x)$ for different time periods as a test of homogeneity. Where we find evidence of homogeneity we then compare the two-transition conditional density with the two-transition iteration of the one-period conditional density as a test of the first-order property. The latter test, which, as detailed below, is based on the Chapman-Kolmogorov equations for an homogenous first-order process, could in principle serve as a test of the joint hypothesis of homogeneity and first-orderedness. However, our Monte Carlo simulations showed that it had virtually no power against the alternative of a non-homogeneous first-order process, and we so adopted the strategy of first testing for homogeneity by comparing estimated $g_\tau(z|x)$ for different time periods and then testing for first-orderedness, conditional on homogeneity, using the test based on the Chapman Kolmogorov equations. Our homogeneity tests are continuous state space equivalents of the homogeneity tests employed by \citet{bickenbach2003} while our tests of first-orderedness are a formalization of the idea behind the comparison of made by \citet{quah1993empirical}, applied to the continuous state space case. Despite it being unknown in practice, for each test, the length of the transition period, $\tau$, is assumed to be correctly specified, which means that a rejection of the null hypothesis in either case could be due to the incorrect choice of $\tau$. For this reason, we perform both tests for a variety of values of $\tau$.

To outline our test of homogeneity, assume that, for a sample of $n$ entities (here countries) the process is observed for three time periods: $t$, $t+\tau$, and $t+2\tau$. Denote the $\tau$-period-ahead density of $z$ conditional on $x$ prevailing at time $t$ as $g_{\tau\text{,}t}(z|x)$. Homogeneity can then be tested by a test of the following $H_0$ against the following $H_1$:
\begin{eqnarray} \nonumber
	H_0 &:& g_{\tau,t}(z|x)=g_{\tau,t+\tau}(z|x) \text{ for all } (z,x)\in\mathbb{R}\times\mathbb{R} \\
	H_1 &:& g_{\tau,t}(z|x)\neq g_{\tau,t+\tau}(z|x) \text{ for some } (z,x)\in\mathbb{R}\times\mathbb{R}.
	\label{test:homogeneity}
\end{eqnarray}
Performing such a test presents two important issues. The first is the appropriate metric for measuring the distance between the estimated values of $g_{\tau\text{,}t}(z|x)$ and $g_{\tau\text{,}t+\tau}(z|x)$. It turns out that the size and power properties the test we propose depend on the metric used. We describe the metrics that we use in Section \ref{sec:measureDivergenceConditionalDensities} and their properties in Section \ref{sec:testProperties}. The second issue is that the asymptotic null distribution of the test statistic for testing \ref{test:homogeneity} is difficult to characterize and so we use a bootstrap procedure, whose details are given in Section \ref{sec:bootstrapProcedure}.

To motivate our test of first-orderedness, observe that, using the notation introduced above, the density of $z$ at time $t+2\tau$ can be written as
\begin{equation}
	\begin{aligned}[b]
		f_{t+2\tau}(z)&=\int_\mathbb{R}g_\tau(z|y)f_{t+\tau}(y)dy \\
		&=\int_\mathbb{R}g_\tau(z|y)\left[\int_\mathbb{R}g_\tau(y|x)f_t(x)dx\right]dy \\
		&=\int_\mathbb{R}\int_\mathbb{R}g_\tau(z|y)g_\tau(y|x)f_t(x)dxdy \\
		&=\int_\mathbb{R}\left[\int_\mathbb{R}g_\tau(z|y)g_\tau(y|x)dy\right]f_t(x)dx.
	\end{aligned}
	\label{eq:distributionDynamicsSecondOrder}
\end{equation}
By analogy to the Equation \eqref{eq:distributionDynamics} above, letting $g_{2\tau,t}(z|x)$ denote the $2\tau$-period-ahead density of $z$ conditional on $x$, the density of $z$ at time $t+2\tau$ can also be written as
\begin{equation}
	f_{t+2\tau}(z)=\int_\mathbb{R}g_{2\tau}(z|x)f_t(x)dx.
	\label{eq:distributionDynamicsSecondOrderII}
\end{equation}
The first-order specification thus holds if
\begin{equation}
	g_{2\tau}(z|x)=\int_\mathbb{R}g_\tau(z|y)g_\tau(y|x)dy,
	\label{eq:distributionDynamicsSecondOrderIII}
\end{equation}
which is just one of the Chapman Kolmogorov equations for the first-order
process.\footnote{Applications of specification tests based on the Chapman-Kolmogorov equations exist in other contexts. See, for example,  \citet{ait1996testing} and \citet{ait2010nonparametric} for applications of the approach to diffusion processes.}

The first-order property can then be tested by a test of the following $H_0$ against the following $H_1$:
\begin{eqnarray}\nonumber
	H_0 &:& g_{2\tau}(z|x)    =    \int_\mathbb{R}g_\tau(z|y)g_\tau(y|x)dy \text{ for all }(z,x)\in \mathbb{R}\times\mathbb{R}\\
	H_1 &:& g_{2\tau}(z|x) \neq \int_\mathbb{R}g_\tau(z|y)g_\tau(y|x)dy \text{ for some }(z,x)\in\mathbb{R}\times\mathbb{R}.
	\label{test:ChapmanKolmogorov}
\end{eqnarray}

An important caveat, which has implications for the power of our test, is
that, as \citet{cox_miller_stochastic_1977} and \citet{chen2012testing} point out, satisfying the Chapman Kolmogorov equations is implied by homogeneity and
first-orderedness of the process but satisfying the Chapman Kolmogorov
equations is not sufficient to imply those properties hold. The issue of
sensitivity of the size and power properties the test we propose to the
metric used to measure the distance between the estimated values of
$g_{2\tau,t}(z|x)$ and $\int_\mathbb{R}g_\tau(z|y)g_\tau(y|x)dy$ arises here
as it does for the homogeneity test above. Again we use a bootstrap procedure
to estimate null distribution of the test statistic. Details are given in Section \ref{sec:bootstrapProcedureOrderProcess}.

\section{ Estimation and testing issues \label{sec:testingMarkovProperties}}

In Section \ref{sec:estimationConditionalDensities} we outline our approach to estimating the conditional densities that are the transition kernels of the process describing the evolution of the cross-country distribution of per worker GDP. All of our hypothesis tests require the computation of the distance between estimated densities and Section \ref{sec:measureDivergenceConditionalDensities} provides the metrics that we use to do so. Sections \ref{sec:bootstrapProcedure} and \ref{sec:bootstrapProcedureOrderProcess} discuss the bootstrap procedures that we use to assess the statistical significance of the estimated distances between distributions in order to test the homogeneity and first-orderedness hypotheses respectively. Section \ref{sec:testProperties} briefly summarizes the Monte Carlo experiments presented in Appendix \ref{app:MonteCarloExperiments} that are designed to explore the properties of our tests.

\subsection{Estimation of conditional densities}
\label{sec:estimationConditionalDensities}

To estimate the conditional densities $g_{\tau\text{,}t}(z|x)$, $g_{\tau\text{,}t+\tau}(z|x)$, $g_\tau(z|y)$, and $g_{2\tau}(z|x)$ needed to
		conduct our hypothesis tests we first estimate the appropriate joint
		distribution which is then used to compute an estimate of the implied
		conditional distribution. For example, given data on per capita income
		$(y_{i,t+2\tau},\,y_{i,t+\tau},\,y_{i,t})$ where $i$ indexes the
		observational unit, $i=1,\ldots,n$, the conditional density
		$g_\tau(y_{t+\tau}|\,y_t)$ is estimated by first estimating the joint density
		$j_\tau(y_{t+\tau},\,y_t)$ using an adaptive kernel density
		estimator \citep[p101]{silverman86}.
This estimator begins with a pilot estimate of the joint density
		$\tilde{j_\tau}(y,x)$ which is computed using a standard
		kernel density estimator with a normal kernel and the optimal normal
		bandwidth \citep[p101]{silverman86}. The adaptive kernel density estimator
		of $j_\tau(y,x)$ is then computed as
\begin{equation}
	\widehat{j}_\tau(y,x)=\frac{1}{nh^2}\sum_{i=1}^n\frac{1}{\lambda_i^2}K\left(%
	\frac{(y,x)-(y_{i,t+\tau},\,y_{i,t})}{h\lambda_i}\right),
	\label{eq:adaptiveKernelEstimator}
\end{equation}
where $K(\cdot)$ is the normal kernel, $h$ is the optimal normal bandwidth,
$\lambda_i=\left(\frac{\tilde{j_\tau}(y_{i,t+\tau},\,y_{i,t})}{g}\right)^{-\alpha}$, $g$ is the geometric mean of the non-zero
$\tilde{j_\tau}(y_{i,t+\tau},\,y_{i,t})$, and $\alpha$ is a
parameter that we set to $0.5$ following the advice of \citet[p103]{silverman86}.\footnote{Setting $\alpha=0$ reduces the adaptive estimator to the standard kernel density estimator.}
The adaptive kernel estimator performs better than the standard kernel estimator because it uses a larger bandwidth where the data are sparse and a smaller one where the data are dense.

Given the estimated joint density, $\widehat{j}_\tau(y_{t+\tau},\,y_t)$, the implied estimate of the marginal density of $y_t$ is calculated as $\widehat{m}(y_t)=\int_\mathbb{R}$
		$\widehat{j}_\tau(y_{t+\tau},\,y_t)dy_{t+\tau}$, and the estimated
		conditional density is computed as $\widehat{g}_\tau(y_{t+\tau}|\,y_t)=\frac{\widehat{j}_\tau(y_{t+\tau},\,y_t)}{\widehat{m}(y_t)}$ using the definition of the conditional density (if it exists) as the ratio of the joint density to the marginal density, where the latter is assumed to be nonzero.
		Similarly, $g_{2\tau}(z|x)$ is estimated as
		$\widehat{g}_{2\tau}(y_{t+2\tau}|y_t)=\frac{\widehat{j}_{2\tau}(y_{t+2\tau},\,y_t)}{\widehat{m}(y_t)}$, where  $\widehat{j}_{2\tau}(y_{t+2\tau},\,y_t)$ is the estimated joint density of
		$y_{t+2\tau}$ and $y_t$, and $\widehat{m}(y_t)=\int_\mathbb{R}$
		$\widehat{j}_{2\tau}(y_{t+2\tau},\,y_t)dy_{t+2\tau}$ is the implied estimate
		of the marginal density of $y_t$.

\subsection{Measures of the divergence between conditional densities}
\label{sec:measureDivergenceConditionalDensities}

As suggested also by \citet{anderson1994two} and \citet{li1996nonparametric}, the tests of the hypotheses given in Equations \eqref{test:homogeneity} and \eqref{test:ChapmanKolmogorov} can be formulated using a measure of the divergence between two conditional distributions. For the two distributions be $f_1(y|x)$ and $f_2(y|x)$, a generic measure of the divergence between them can be constructed using a real non-negative valued function
\begin{equation}
	D=D(f_1(y|x),f_2(y|x),\,\omega(x)),
	\label{eq:divergenceMeasure}
\end{equation}
where $\omega(x)$ is a non-negative weighting function. \citet{pardo2018statistical} provides a deep discussion of different measures of divergence (see also \citealp{marron1995visual}) and among those, the most used are those based on the $L^p$ norms
\begin{equation}
	D^{NORM}_p=\left[\int_\mathbb{R}\int_\mathbb{R}|f_1(y|x)-f_2(y|x)|^p%
	\omega(x)dydx\right]^{\frac{1}{p}},
	\label{eq:LpNorm}
\end{equation}
with $p \in\{1,\,2,\,\infty\}$ corresponding to the $L^1,\,L^2,\,$and $L^\infty$ norms respectively.
Additionally, we employ a measure of
divergence based on the Hellinger distance\footnote{\cite{su2008nonparametric} use the Hellinger distance in the construction of a test of conditional independence.}
\begin{equation}
	D^H=\left\{\frac{1}{2}\int_\mathbb{R}\int_\mathbb{R}\left[\sqrt{f_1(y|x)}-%
	\sqrt{f_2(y|x)}\right]^2\omega(x)dydx\right\}^{\frac{1}{2}}.
	\label{eq:HillengerDistance}
\end{equation}
All of these metrics include the $\omega(x)$ function which, following \cite[p3134]{ait2010nonparametric}, ``is a weight function selected to reduce the influences of the unreliable estimates in the sparse region''. That is, we adjust the distance measures that we use so that greater weight is given to regions where our (kernel density) estimators are more precise due to the presence of more data points.

All of our tests involve using one or more of these metrics to measure	the distance between estimates of $g_{\tau\text{,}t}(z|x)$ and $g_{\tau\text{,}t+\tau}(z|x)$ in the case of homogeneity tests or estimates of \ $g_{2\tau}(z|x)$ and $\int_\mathbb{R}g_\tau(z|y)g_\tau(y|x)dy$ in the case of tests of the first-order property. So, for example, in the case of homogeneity tests, the test statistic using the Hellinger distance is computed as
\begin{equation}
	\widehat{D}^H=\left\{\frac{1}{2}\int_\mathbb{R}\int_\mathbb{R}\left[\sqrt{%
		\widehat{g}_{\tau\text{,}t}(y|x)}-\sqrt{\widehat{g}_{\tau\text{,}t+%
			\tau}(y|x)}\text{
	}\right]^2\widehat{m}(x)dydx\right\}^{\frac{1}{2}}.
		\label{eq:testStatisticsHellingerDistance}
\end{equation}
where $\widehat{g}_{\tau\text{,}t}(y|x)$ is the estimated one-period
transition kernel based on $(y_{t+\tau},y_t)$,
$\widehat{g}_{\tau\text{,}t+\tau}(y|x)$ is the estimated one-period
transition kernel based on $(y_{t+2\tau},y_{t+\tau}),$ and $\widehat{m}(x)$
is the estimated marginal density of $y_t$. As mentioned above, in both
cases, we use the bootstrap approach of \citet{efrom1993introduction} to
evaluate the statistical significance of the measured distances between the
estimated distributions.

\subsection{Bootstrap procedure for testing time homogeneity}
\label{sec:bootstrapProcedure}

We use the triple of data on per capita GDP
	$(y_{i,t+2\tau},\,y_{i,t+\tau},\,y_{i,t})$, $i=1,\ldots,n$, to calculate
	the estimated conditional distributions
	$\widehat{g}_{\tau\text{,}t}(y_{t+\tau}|y_t)$,
	$\widehat{g}_{\tau\text{,}t+\tau}(y_{t+2\tau}|y_{i,t+\tau})$, and
	$\widehat{m}(y_t)$, the estimated marginal distribution of $\,y_t$, using the approach described above.\footnote{Following \citet{hall2001bootstrapping}, here and in the tests of the order of the process, we use the bandwidth estimated when estimating the conditional distributions for the bootstrap replications rather than re-estimating the
	bandwidth for each replication.}
	These estimates are used to calculate the measure of divergence between the two estimated conditional distributions $\widehat{D}(OBS)=D(\widehat{g}_{\tau\text{,}t}(y_{t+\tau}|y_t),\widehat{g}_{\tau\text{,}t+\tau}(y_{t+2\tau}|y_{i,t+\tau}),\widehat{m}(y_t))$, where the exact expression for $D$ depends on the distance metric used. For the $b^{\text{th}}$ bootstrap replication, we pool the observed one-period transition data $(y_{i,t+2\tau},y_{i,t+\tau})$ and $(y_{i,t+\tau},\,y_{i,t})$ data and draw two artificial samples of pairs of observations $(y^b_{i,t+2\tau},y^b_{i,t+\tau})$ and $(y^b_{i,t+\tau},\,y^b_{i,t})$, each of size $n$, by sampling with replacement from the pooled data to generate two samples respecting the null hypothesis $H_0$.
	Using the approach described above, we then estimate the conditional distributions producing $\widehat{g}^b_{\tau\text{,}t}(y_{t+\tau}|y_t),\,\widehat{g}^b_{\tau\text{,}t+\tau}(y_{t+2\tau}|y_{i,t+\tau})$, and $\,\,\widehat{m}^b(y_t)$, which are used to calculate the implied measure of divergence between the two estimated conditional distributions, $\widehat{D}(b)=D(\widehat{g}^b_{\tau\text{,}t}(y_{t+\tau}|y_t),%
	\widehat{g}^b_{\tau\text{,}t+\tau}(y_{t+2\tau}|y_{i,t+\tau}),%
	\widehat{m}^b(y_t))$ for this replication. We repeat the replication process $B$ times and calculate the achieved significance level of the test statistic as $ASL=\sum_{b=1}^BI(\widehat{D}(b)>\widehat{D}(OBS))/B$, where $I(\cdot)$ is an indicator function.

\subsection{Bootstrap procedure for testing the order of process}
\label{sec:bootstrapProcedureOrderProcess}

	   We use the triple of data on per capita GDP
		$(y_{i,t+2\tau},\,y_{i,t+\tau},\,y_{i,t})$, $i=1,\ldots,n$, to \ calculate
		the estimated conditional distributions $\widehat{g}_\tau(y_{t+\tau}|\,y_t)$,
		$\widehat{g}_{2\tau}(y_{t+2\tau}|y_t)$, and $\widehat{m}(y_t)$, the estimated
		marginal distribution of $\,y_t$, using the approach described above. These
		estimates are used to calculate the measure of divergence between the two
		estimated distributions of $y_{t+2\tau}$ conditional on $y_t$
		\[
		\widehat{D}(OBS)=D\left(\widehat{g}_{2\tau}(y_{t+2\tau}|y_t),\int_\mathbb{R}\widehat{g}_\tau(y_{t+2\tau}|\,y_{t+\tau})\widehat{g}_\tau(y_{t+\tau}|\,y_t)dy_{t+\tau},\widehat{m}(y_t)\right),
		\] where the exact expression for $D$ depends
		on the distance metric used. For the $b^{\text{th}}$ bootstrap replication,
		we draw an artificial sample of one-period transition data
		$(y^b_{i,t+\tau},y^b_{i,t})$ of size $n$, by sampling with replacement from
		the $(y_{i,t+\tau},\,y_{i,t})$ data and estimate the conditional density
		$\widehat{g}^b_\tau(y_{t+\tau}|\,y_t)$ and the marginal density
		$,\widehat{m}^b(y_t)$ using the approach described above. We then generate
		$y^b_{i,t+2\tau}$ based on $y^b_{i,t+\tau}$ and $\widehat{g}_\tau(y_{t+\tau}|\,y_t)$ by ``rejection sampling'' \citep[p47]{robert2004monte} to yield an artificial triple
		$(y^b_{i,t+2\tau},\,y^b_{i,t+\tau},\,y^b_{i,t})$ respecting the null hypothesis $H_0$.
		Finally, we estimate
		$g_{2\tau}(y_{t+2\tau}|y_t)$ giving		$\widehat{g}^b_{2\tau}(y_{t+2\tau}|\,y_t)$ using the approach described above
		and then use the $\widehat{g}^b_{2\tau}(y_{t+2\tau}|\,y_t)$,
		$\widehat{g}^b_\tau(y_{t+\tau}|\,y_t)$, and $\widehat{m}^b(y_t)$ to calculate
		the measure of divergence between the two estimated conditional distributions
		\[
		\widehat{D}(b)=D\left(\widehat{g}^b_{2\tau}(y_{t+2\tau}|y_t),\int_\mathbb{R} \widehat{g}^b_\tau(y_{t+2\tau}|\,y_{t+\tau})\widehat{g}^b_\tau(y_{t+\tau}|\,y_t)dy_{t+\tau},\widehat{m}^b(y_t)\right)
		\] for this replication.
		We repeat the replication process B times and calculate the achieved significance level of the test statistic as $ASL=\sum_{b=1}^BI(\widehat{D}(b)>\widehat{D}(OBS))/B$,
		where $I(\cdot)$ is an indicator function.

\subsection{Test properties}
		\label{sec:testProperties}

		In Appendix \ref{app:MonteCarloExperiments} we report the results of some Monte Carlo experiments designed to explore the properties of our proposed testing procedures, in particular the effect of the different measures of divergence between distributions, on the empirical size and power of our tests. In general, those experiments lead us to conclude that, for both the tests of time homogeneity and of the order of the process, the $L^1$ and $H$ distance measures generally perform well in terms of the empirical size of the tests that we propose even when the sample size is as small as 100. The only exception to this result seems to be that $L^2$ and $L^\infty$ exhibit better empirical size properties for the tests of the order of the process when the persistence in the process is high. $L^1$ and $H$ also perform better than the other distance measures for the homogeneity tests in terms of the empirical power of the tests, although the power is low for all distance measures when the sample size and/or the deviation from the null hypothesis is small. For the empirical power of the tests of the order of the process, $L^1$,  $L^2$, and $H$ generally perform be than $L^\infty$, although none of the divergence measures produces high power for small deviations from the null hypothesis, even in large samples.

In the next section we present an application of the tests to PWT 10.0 data on GDP per worker for a panel of 102 countries over the period 1970-2019 before offering some conclusions. To save space, we present only the results of the homogeneity tests using the $L^1$ and $H$ distance measures, those having the better size and power properties according to our Monte Carlo experiments. The complete set of homogeneity test results using all four distance measures is provided in Appendix \ref{app:testingTimeHomogeneity}.

\section{The distribution dynamics of GDP per worker \label{sec:applicationPWT}}

In this section, we implement the tests described above using data from PWT 10.0 \citep{feenstra2015} for the balanced sample of 102 countries over the 1970-2019 period described in Section \ref{sec:datamotivation}.
Before using the tests described above to investigate the stability or lack thereof in the dynamics of the cross-country distribution of per capita GDP, we must first face an important unknown in the specification of process governing the dynamics. Specifically, we do not know the transition length i.e the parameter $\tau$ in the exposition of the tests above. Our approach to this issue, in both the time homogeneity tests, and the tests of the first-orderedness of the process, is to conduct the tests for values of $\tau$ of 5, 10, 15, 20 and 25 years. That is, in all cases, our tests are conditional on having the correct value of $\tau$ but, as the true value of that parameter is unknown, we perform the tests for a range of plausible values.

To enable use of the entire span of the dataset, we use a one-year overlap between the last and next-to-last transition periods for the 5-, 10-, and 25-year transitions. This means that: for the 5-year transitions the last period is 2014 to 2019 and the next-to-last is 2000 to 2015; for the 10-year transitions the last is 2009 to 2019 and the next-to-last is 2000 to 2010; and, for the 25-year transitions, the first is 1970 to 1995 and the second is 1994 to 2019. Additionaly, for the 15-year and 20-year transitions we consider two different starting years: for the 15-year transitions, 1970 and 1974; and, for the 20-year transitions, 1970 and 1979. In each case, the second, later start year means that 2019 is the end year of the last transition used.

\subsection{Testing the time homogeneity \label{sec:timeHomogeneity}}

\subsubsection{5-year transitions \label{sec:5YearSubPeriods}}

\begin{figure}[!htbp]
\begin{subfigure}{0.49\textwidth}
	\includegraphics[width=1.\textwidth]{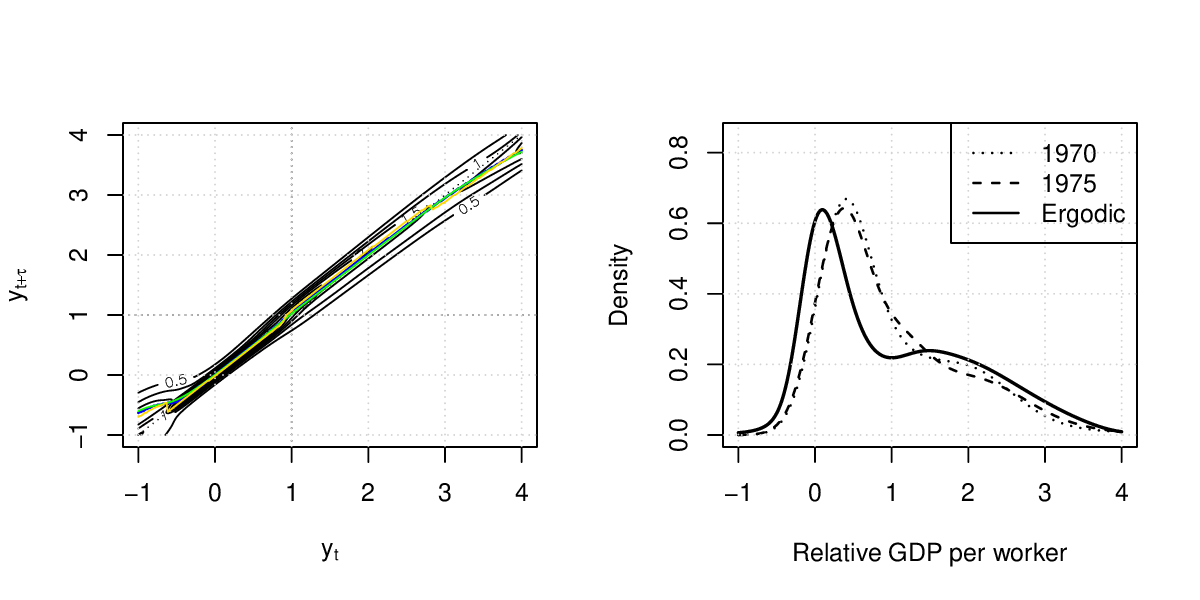}
	\caption{Period 1970-1975}
	\label{fig:stochasticKernelErgodicDistr_1970_1975bandwidth_optimalalpha_0}
\end{subfigure}
\begin{subfigure}{0.49\textwidth}
	\centering
	\includegraphics[width=1.\textwidth]{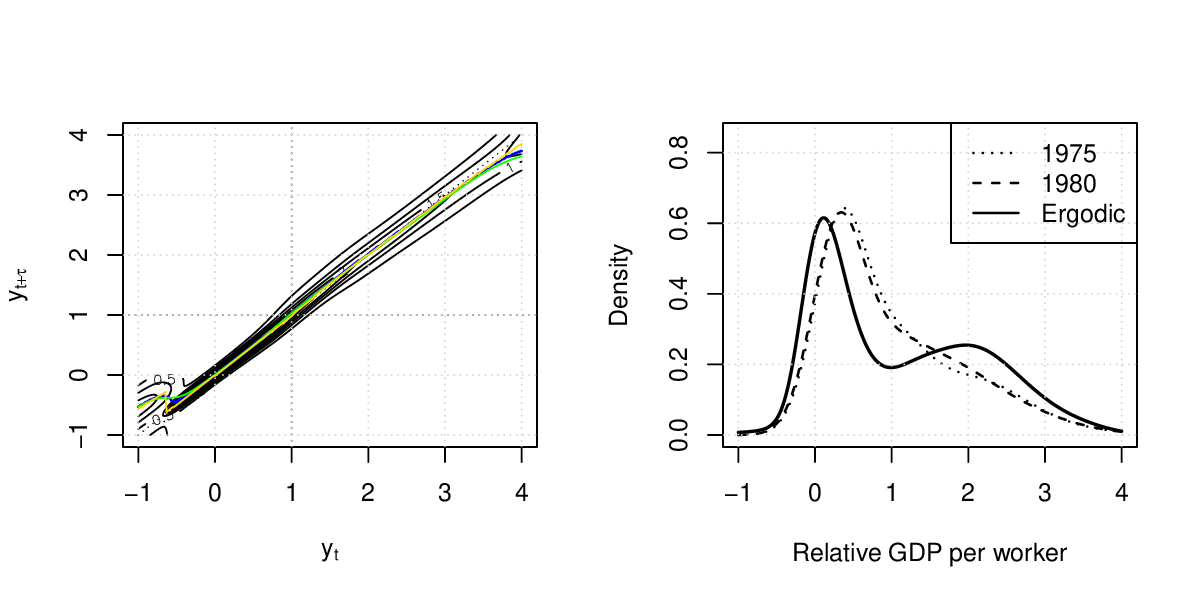}
	\caption{Period 1975-1980}
	\label{fig:stochasticKernelErgodicDistr_1975_1980bandwidth_optimalalpha_0}
\end{subfigure}
\begin{subfigure}{0.49\textwidth}
	\centering
	\includegraphics[width=1.\textwidth]{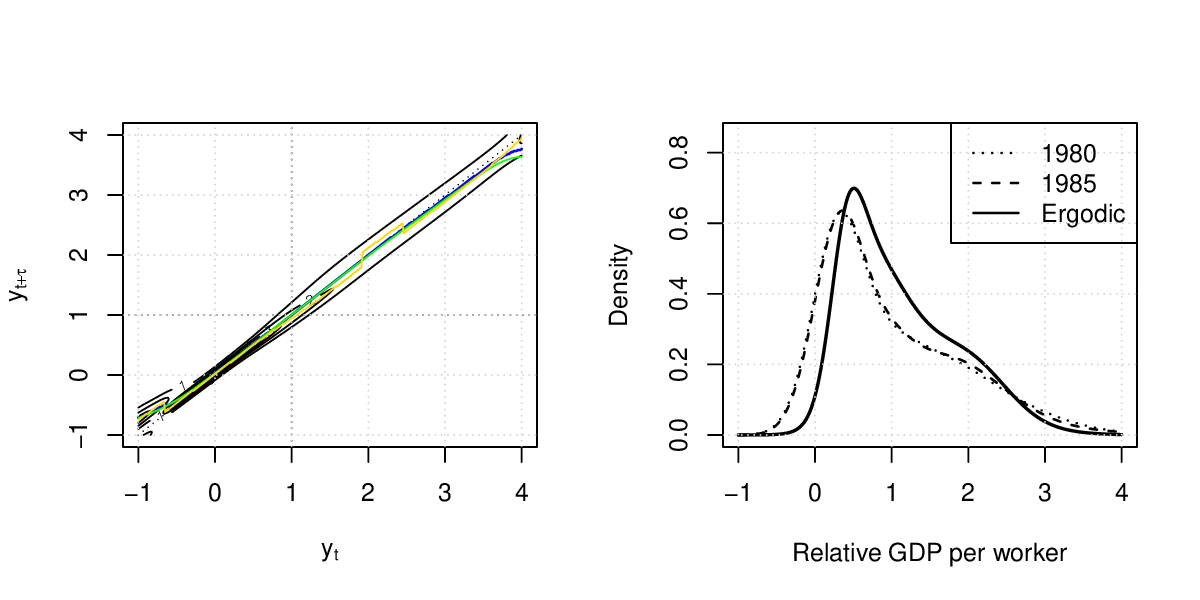}
	\caption{Period 1980-1985}
	\label{fig:stochasticKernelErgodicDistr_1980_1985bandwidth_optimalalpha_0}
\end{subfigure}
\begin{subfigure}{0.49\textwidth}
	\centering
	\includegraphics[width=1.\textwidth]{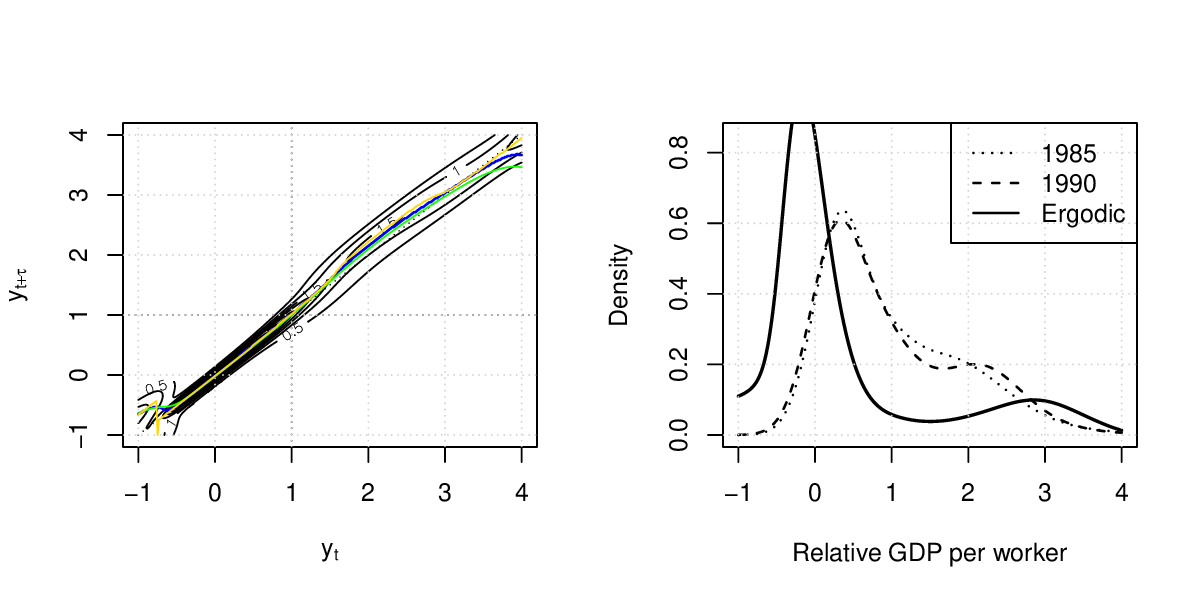}
	\caption{Period 1985-1990}
	\label{fig:stochasticKernelErgodicDistr_1985_1990bandwidth_optimalalpha_0}
\end{subfigure}
\begin{subfigure}{0.49\textwidth}
	\centering
	\includegraphics[width=\textwidth]{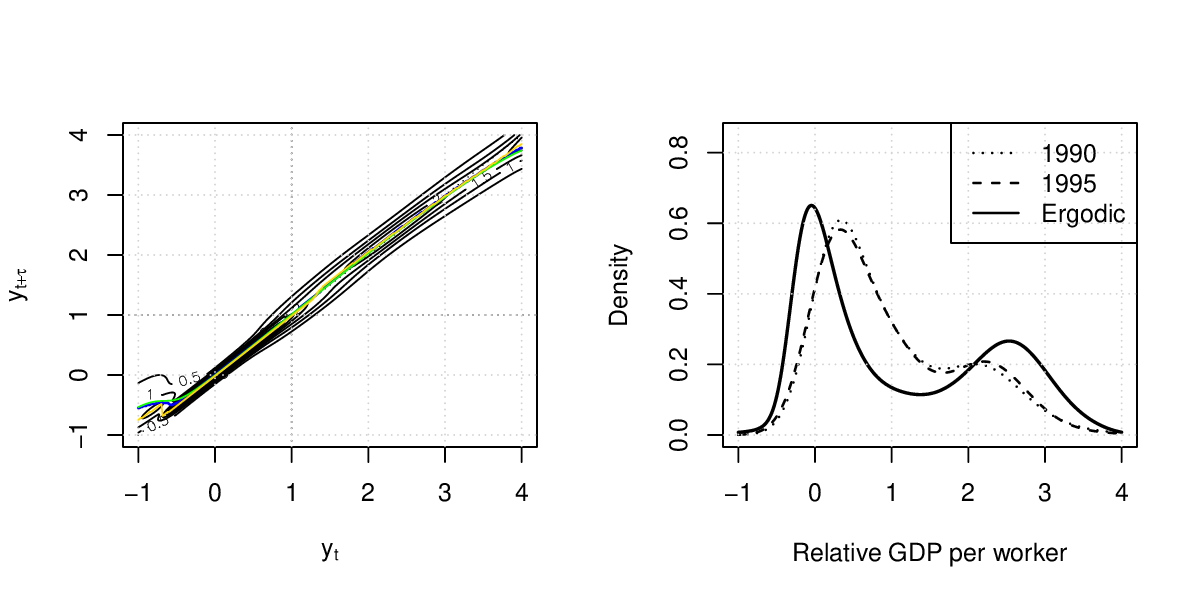}
	\caption{Period 1990-1995}
	\label{fig:stochasticKernelErgodicDistr_1990_1995bandwidth_optimalalpha_0}
\end{subfigure}
\begin{subfigure}{0.49\textwidth}
	\centering
	\includegraphics[width=\textwidth]{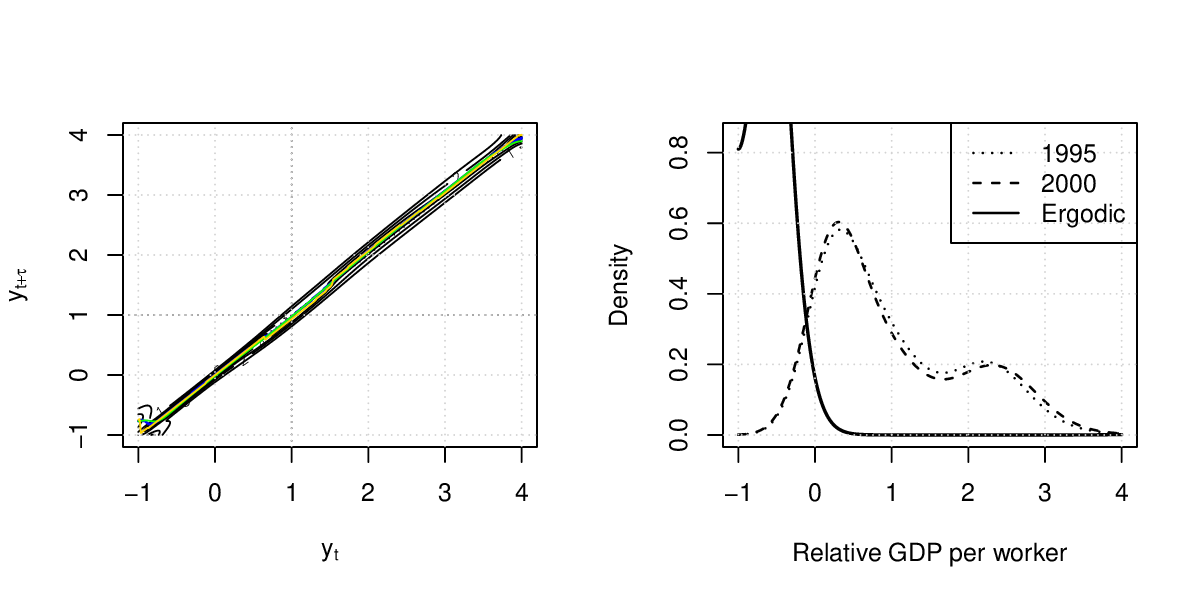}
	\caption{Period 1995-2000}
	\label{fig:stochasticKernelErgodicDistr_1995_2000bandwidth_optimalalpha_0}
\end{subfigure}
\begin{subfigure}{0.49\textwidth}
	\centering
	\includegraphics[width=\textwidth]{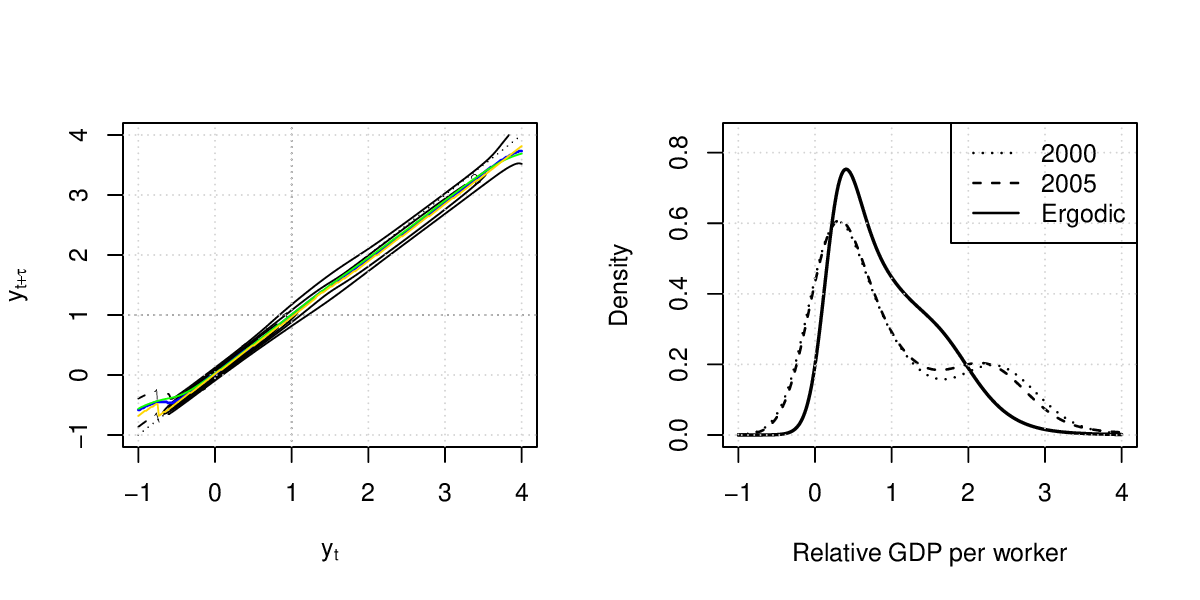}
	\caption{Period 2000-2005}
	\label{fig:stochasticKernelErgodicDistr_2000_2005bandwidth_optimalalpha_0}
\end{subfigure}
\begin{subfigure}{0.49\textwidth}
	\centering
	\includegraphics[width=\textwidth]{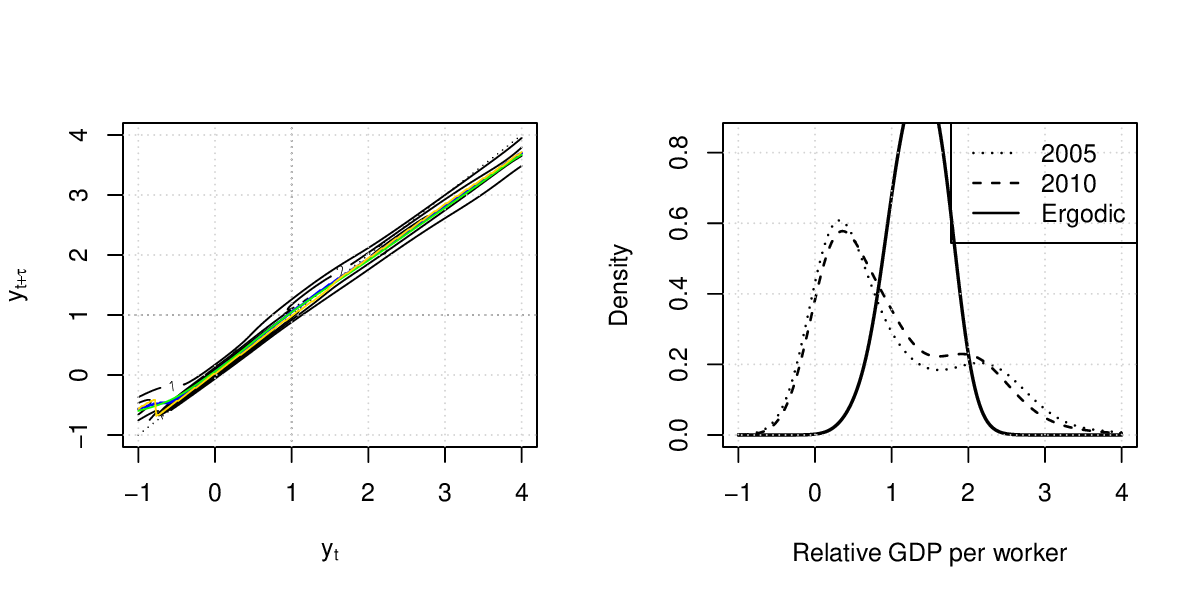}
	\caption{Period 2005-2010}
	\label{fig:stochasticKernelErgodicDistr_2005_2010bandwidth_optimalalpha_0}
\end{subfigure}
\begin{subfigure}{0.49\textwidth}
	\centering
	\includegraphics[width=\textwidth]{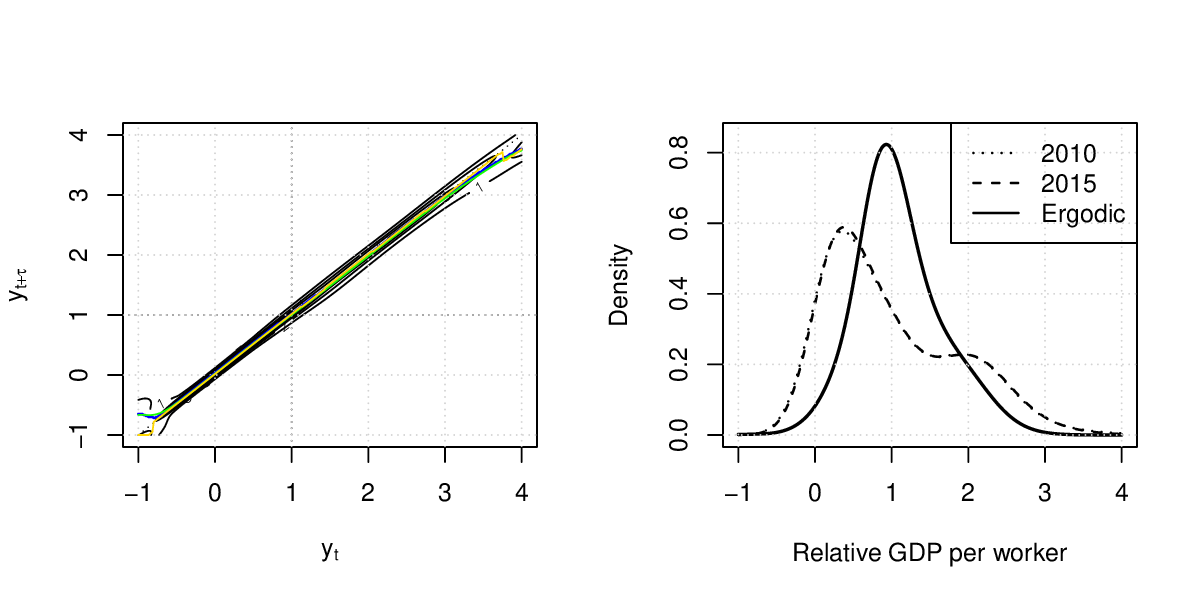}
	\caption{Period 2010-2015}
	\label{fig:stochasticKernelErgodicDistr_2010_2015bandwidth_optimalalpha_0}
\end{subfigure}
\begin{subfigure}{0.49\textwidth}
	\centering
	\includegraphics[width=\textwidth]{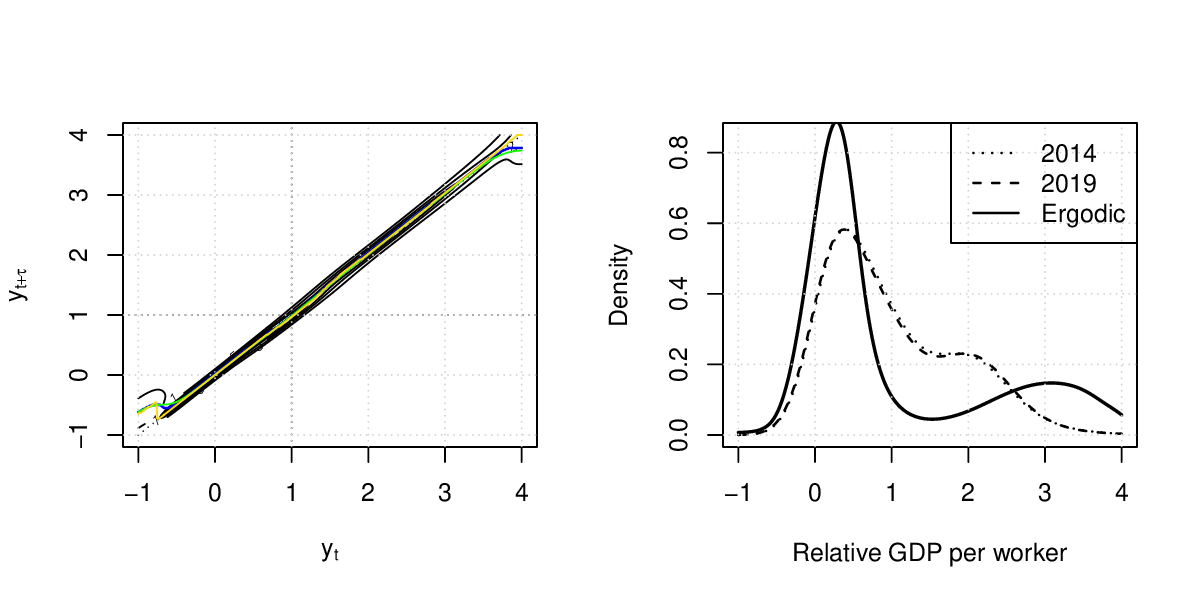}
	\caption{Period 2014-2019}
	\label{fig:stochasticKernelErgodicDistr_2014_2019bandwidth_optimalalpha_0}
\end{subfigure}
\caption{For each 5-year period, the first panel shows the estimated transition kernel of the cross country distribution of per capita GDP and the second shows the estimated density for the first and last year as well as the ergodic density implied by the estimated transition kernel. Estimation used a adaptive kernel estimator with a Gaussian kernel and the optimal Normal bandwidth \citep[p101]{silverman86}. The range of estimation is $[-1,4]$ in a regular grid of $100 \times 100$ evaluation points. The estimated transition kernel is calculated from the estimated joint distribution of the first and last years' per capita GDP levels as explained in the text. Authors' calculations using data from \citet{feenstra2015}.}
\label{fig:5YearTransitions}
\end{figure}

The panels in Figure \ref{fig:5YearTransitions} show, for each 5-year transition, the estimated transition kernel, the estimated densities in the initial and final year, and the ergodic density implied by the estimated transition kernel. While most of the densities estimated for particular years exhibit the well-known twin-peaked shape, the change in the distribution dynamics implied by the results of the homogeneity tests is evident in the ergodic distributions. Those computed for transitions up to 1995 are generally fairly similar, exhibiting twin-peakedness or almost twin-peakedness, akin to that found for the densities estimated for particular years. However, the ergodic distributions implied by the estimated transition kernels for 1995-2000, 2005-2010, and 2010-2015 differ markedly from those for the other periods in that they are all strongly single-peaked. This suggests that very different transition dynamics were in play during those periods. We formalize these perceived changes in the distribution dynamics with the time homogeneity tests (supposing that the transition period is 5 years) presented in Table \ref{tab:testTimeHomogeneity5years} and in Table \ref{tab:testTimeHomogeneity5years_app} in Appendix \ref{app:testingTimeHomogeneity}.

\begin{table}[!htbp]
	\centering
	\hspace{-1cm}
	\scriptsize{
	\begin{tabular}{rHrrrrrrrrr}
		\hline
		\hline
		\textbf{Test based on $L^1$} & 1970-1975 & 1975-1980 & 1980-1985 & 1985-1990 & 1990-1995 & 1995-2000 & 2000-2005 & 2005-2010 & 2010-2015 & 2014-2019 \\
		\hline
		1970-1975 &  & 0.92 & 0.40 & 0.44 & 0.96 & \textbf{0.00} & 0.17 & \textbf{0.01} & \textbf{0.00} & \textbf{0.00} \\
		1975-1980 &  &  & 0.18 & 0.64 & 0.85 & \textbf{0.00 }& 0.19 & \textbf{0.00} & \textbf{0.00} & \textbf{0.01} \\
		1980-1985 &  &  &  & 0.06 & 0.34 & \textbf{0.00} & 0.65 & 0.10 & \textbf{0.01} & \textbf{0.03} \\
		1985-1990 &  &  &  &  & 0.40 & \textbf{0.00} & \textbf{0.01 }& \textbf{0.00} & \textbf{0.00 }& \textbf{0.00} \\
		1990-1995 &  &  &  &  &  & \textbf{0.02} & 0.21 & \textbf{0.00} & \textbf{0.00 }&\textbf{ 0.01} \\
		1995-2000 &  &  &  &  &  &  & \textbf{0.00} & \textbf{0.00} & \textbf{0.00} & 0.33 \\
		2000-2005 &  &  &  &  &  &  &  & 0.10 & \textbf{0.02} & \textbf{0.04} \\
		2005-2010 &  &  &  &  &  &  &  &  & \textbf{0.01} & \textbf{0.00 }\\
		2010-2015 &  &  &  &  &  &  &  &  &  & 0.19 \\
		\hline
	\\
		\textbf{Test based on $H$} & 1970-1975 & 1975-1980 & 1980-1985 & 1985-1990 & 1990-1995 & 1995-2000 & 2000-2005 & 2005-2010 & 2010-2015 & 2014-2019 \\
		\hline
		1970-1975 &  & 0.86 & 0.35 & 0.29 & 0.99 & \textbf{0.00} & 0.46 &\textbf{ 0.02} & \textbf{0.01} & \textbf{0.02} \\
		1975-1980 &  &  & 0.14 & 0.71 & 0.89 & \textbf{0.00} & 0.31 & \textbf{0.00 }& \textbf{0.00 }& \textbf{0.04} \\
		1980-1985 &  &  &  & 0.09 & 0.24 & \textbf{0.01} & 0.31 & 0.11 & 0.10 & 0.12 \\
		1985-1990 &  &  &  &  & 0.45 & \textbf{0.01 }& \textbf{0.03} & \textbf{0.00 }& \textbf{0.01 }& \textbf{0.03} \\
		1990-1995 &  &  &  &  &  & \textbf{0.01} & 0.38 & \textbf{0.00} & \textbf{0.01 }& 0.16 \\
		1995-2000 &  &  &  &  &  &  & \textbf{0.00 }& \textbf{0.00 }& \textbf{0.00} & 0.22 \\
		2000-2005 &  &  &  &  &  &  &  & 0.06 & 0.07 & 0.07 \\
		2005-2010 &  &  &  &  &  &  &  &  & \textbf{0.01} & \textbf{0.00} \\
		2010-2015 &  &  &  &  &  &  &  &  &  & 0.51 \\
			\hline
			\\
		Number of obs & 102 & 102 & 102 & 102 & 102 & 102 & 102 & 102 & 102  & 102 \\
		\hline
		\hline
	\end{tabular}
}
	\caption{ASL values computed for the null hypothesis of time-homogeneity for the indicated 5-year sub-periods using the procedure described in the text. Authors' calculations using data from \citet{feenstra2015}.}
	\label{tab:testTimeHomogeneity5years}
\end{table}

Table \ref{tab:testTimeHomogeneity5years} presents the ASL values for the bootstrap procedure for the time homogeneity tests that use the $L^1$ and $H$ distance measures, those having the better size and power properties according to our Monte Carlo experiments, when the transition period is taken to be 5 years.\footnote{Table \ref{tab:testTimeHomogeneity5years_app} in Appendix \ref{app:testingTimeHomogeneity} gives the ASL values for these tests each of the four distance measures described above when the transition period is taken to be 5 years. Subsequent tables in Appendix \ref{app:testingTimeHomogeneity} present the full set of ASL values for the 10, 15, 20, and 25 year transition lengths respectively.} The ASL values less that 0.05 are shown in bold and are taken as indicative of a rejection of the null hypothesis of a time homogeneous process across the indicated sub-periods. These tests fail to reject the null hypothesis when periods ending prior to 1995 are compared with each other. Comparisons of periods ending prior to 1995 with those beginning in 1995 or later, yield many more rejections, except in the cases of 1970-1975, 1975-1980, 1980-1985, 1990-1995, compared with 2000-2005, the cases of 1995-2000 and 2010-2015 compared to 2014-2019. These results are consistent with a process that is time homogeneous over the 1970 to 1995 period, the first half of our sample, but strongly suggest a break in the process in the mid to late 1990s and the possibility of time homogeneity towards the end of the sample although the evidence of a return to a homogeneous process is much weaker than that of time homogeneity prior to 1995. To the extent that the implied ergodic densities are reliable summary statistics for the distribution dynamics, these results are entirely consistent with findings in Figure \ref{fig:5YearTransitions} of ergodic densities for 1995-2000, 2005-2010, and 2010-2015 that are very different from those in the other 5-year periods.

\subsubsection{10-year transitions	\label{sec:10YearSubPeriods}}

\begin{figure}[!htbp]
\begin{subfigure}{0.49\textwidth}
	\includegraphics[width=1.\textwidth]{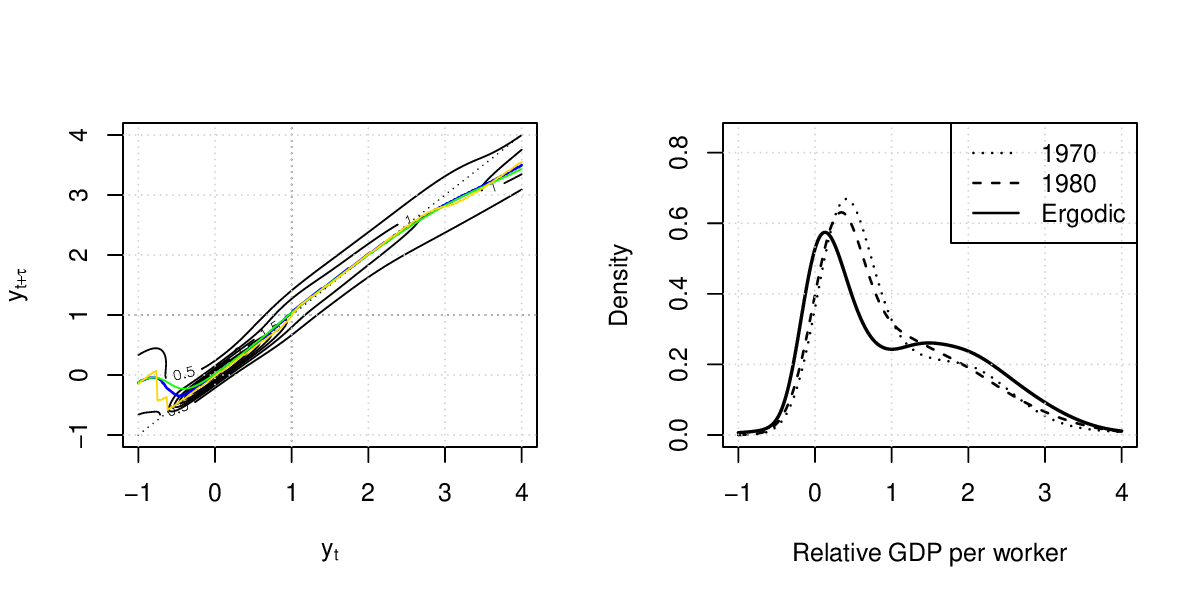}
	\caption{Period 1970-1980}
	\label{fig:stochastickernelergodicdistr19701980bandwidthoptimalalpha0}
\end{subfigure}
\begin{subfigure}{0.49\textwidth}
	\centering
	\includegraphics[width=1.\textwidth]{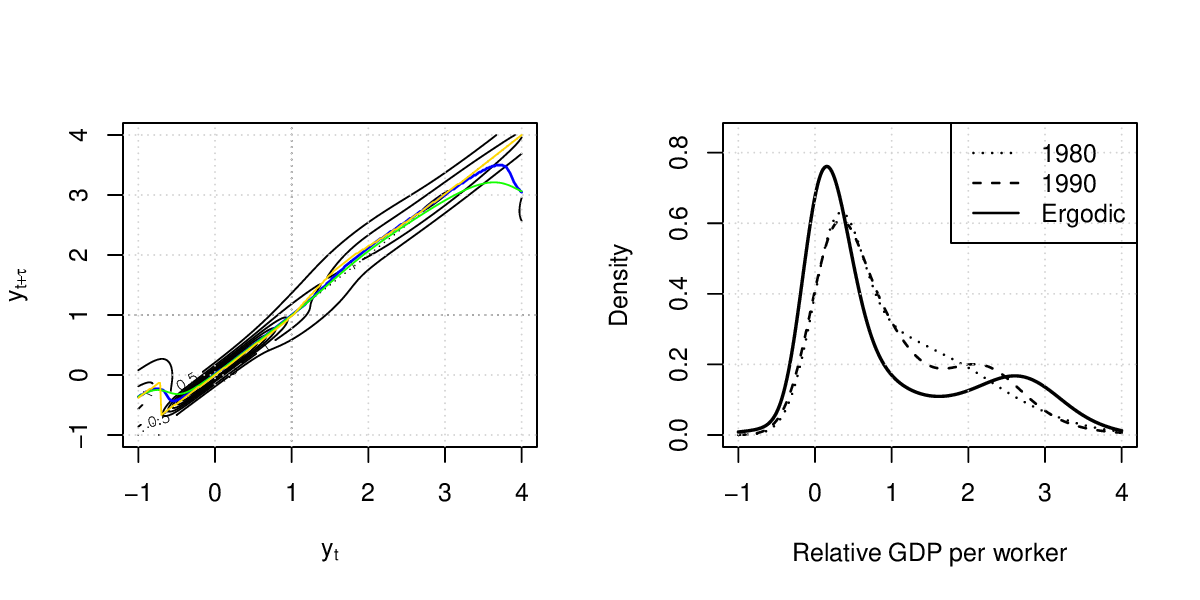}
	\caption{Period 1980-1990}
	\label{fig:stochastickernelergodicdistr19801990bandwidthoptimalalpha0}
\end{subfigure}
\begin{subfigure}{0.49\textwidth}
	\centering
	\includegraphics[width=1.\textwidth]{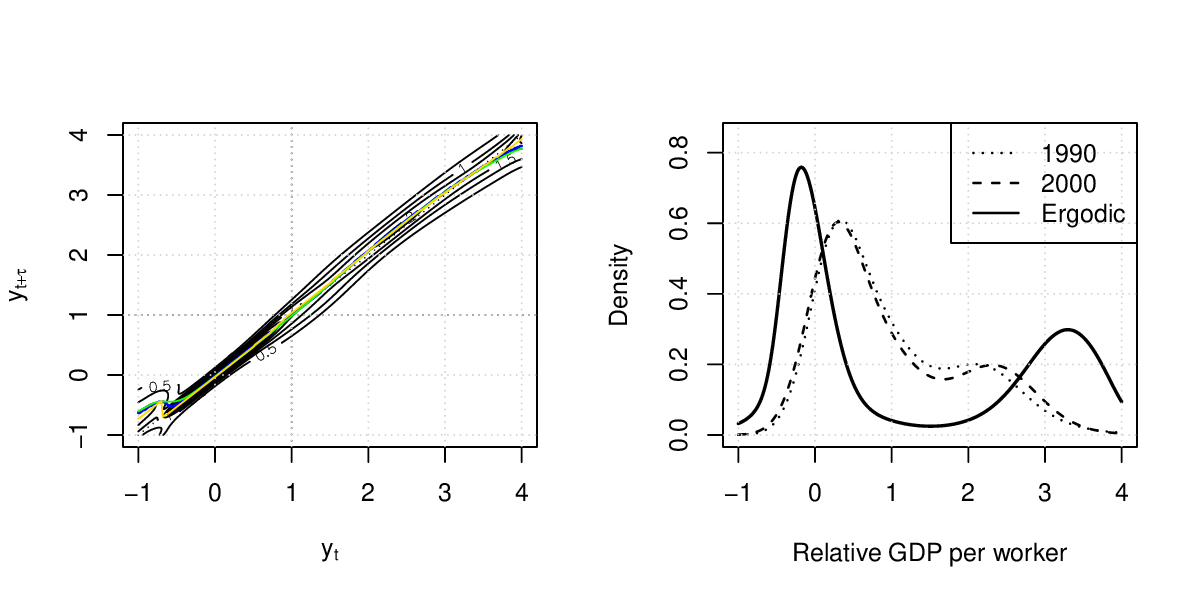}
	\caption{Period 1990-2000}
	\label{fig:stochastickernelergodicdistr19902000bandwidthoptimalalpha0}
\end{subfigure}
\begin{subfigure}{0.49\textwidth}
	\centering
	\includegraphics[width=1.\textwidth]{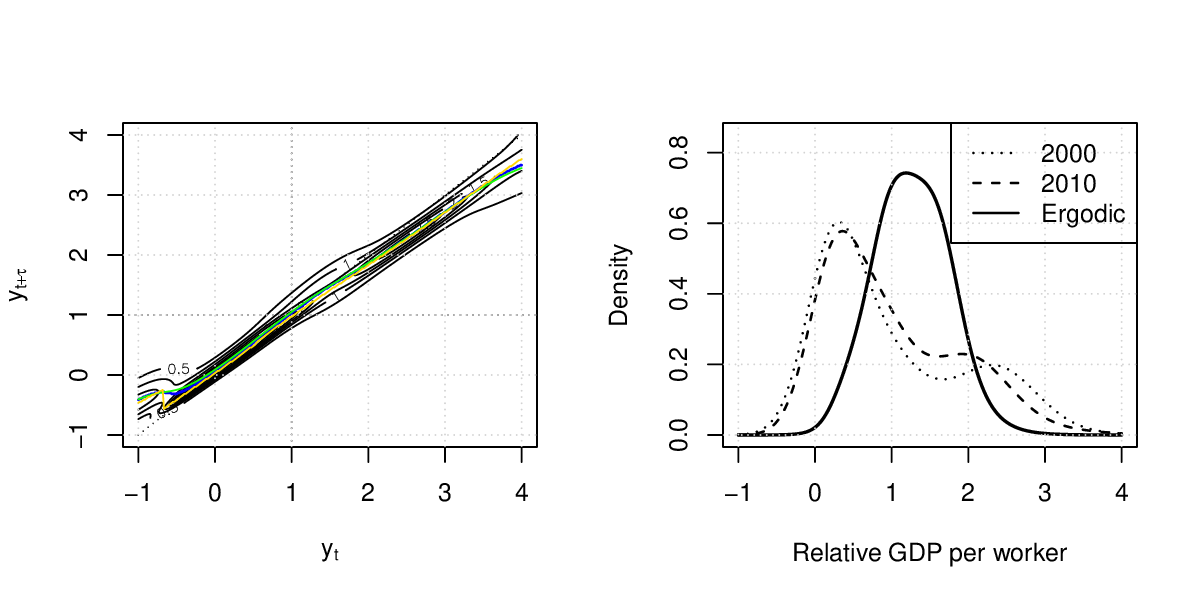}
	\caption{Period 2000-2010}
	\label{fig:stochastickernelergodicdistr20002010bandwidthoptimalalpha0}
\end{subfigure}
\begin{subfigure}{0.49\textwidth}
	\centering
	\includegraphics[width=\textwidth]{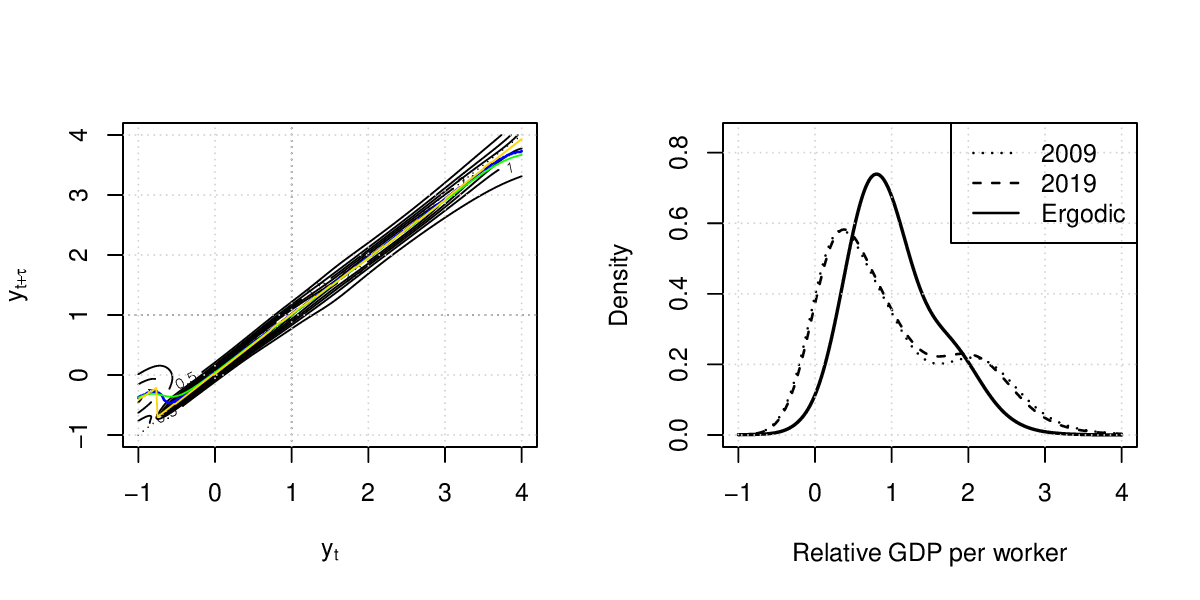}
	\caption{Period 2009-2019}
	\label{fig:stochastickernelergodicdistr20092019bandwidthoptimalalpha0}
\end{subfigure}
\caption{For each 10-year period, the first panel shows the estimated transition kernel of the cross country distribution of per capita GDP and the second shows the estimated density for the first and last year as well as the ergodic density implied by the estimated transition kernel. Estimation used a adaptive kernel estimator with a Gaussian kernel and the optimal Normal bandwidth \citep[p101]{silverman86}. The range of estimation is $[-1,4]$ in a regular grid of $100 \times 100$ evaluation points. The estimated transition kernel is calculated from the estimated joint distribution of the first and last years' per capita GDP levels as explained in the text. Authors' calculations using data from \citet{feenstra2015}.}
\label{fig:10YearTransitions}
\end{figure}

Table \ref{tab:testTimeHomogeneity10YearSubperiods} presents the ASL values for the time homogeneity tests that use the $L^1$ and $H$ distance measures supposing that the transition period is 10 years. The hypothesis of homogeneity cannot be rejected at the 5\% level when 1970-1980 and 1980-1990 are compared, nor can it be rejected when 1980-1990 is compared with 1990-2000, although rejection would occur at the 10\% level. The comparison of 1970-1980 and 1990-2000 yields a rejection at the 5\% level. The other comparisons all result in rejections of homogeneity at the 5\% level. The apparent non-homogeneity of the transition process after 2000 implied by the test results is evident in the plots of the
ergodic distributions shown in Figure \ref{fig:10YearTransitions} as those for 2000-2010 and 2009-2019 are strongly single-peaked, like those for 1995-2000, 2005-2010, and 2010-2015 in the case of 5-year transitions, while those estimated for periods prior to 2000 all exhibit twin-peakedness.

\begin{table}[!htbp]
	\centering
	\begin{tabular}{rHrrrr}
		\hline
		\hline
		\textbf{Test based on $L^1$}  & 1970-1980 & 1980-1990 & 1990-2000 & 2000-2010 & 2009-2019 \\
		\hline
		1970-1980 &  & 0.38 & \textbf{0.00 }& \textbf{0.00 }& \textbf{0.00} \\
		1980-1990 &  &  & 0.06 & \textbf{0.00} & \textbf{0.01} \\
		1990-2000 &  &  &  & \textbf{0.00 }& \textbf{0.01} \\
		2000-2010 &  &  &  &  & \textbf{0.03}\\
		\hline
	\\
	\textbf{Test based on $H$}  & 1970-1980 & 1980-1990 & 1990-2000 & 2000-2010 & 2009-2019 \\
	\hline
	1970-1980 &  & 0.43 & \textbf{0.01} & \textbf{0.01} & \textbf{0.01} \\
	1980-1990 &  &  & 0.07 & \textbf{0.01} & \textbf{0.04} \\
	1990-2000 &  &  &  & \textbf{0.00 }& \textbf{0.03} \\
	2000-2010 &  &  &  &  & \textbf{0.04} \\
	\hline
	\\
	Number of obs & 102 & 102 & 102 & 102 & 102 \\
	\hline
	\hline
	\end{tabular}
	\caption{ASL values computed for the null hypothesis of time-homogeneity for the indicated 10-year sub-periods using the procedure described in the text. Authors' calculations using data from \citet{feenstra2015}.}
	\label{tab:testTimeHomogeneity10YearSubperiods}
\end{table}

\subsubsection{15-year sub-periods 	\label{sec:15YearSubPeriods}}

\begin{figure}[!htbp]
	\begin{subfigure}{0.49\textwidth}
		\centering
		\includegraphics[width=\linewidth]{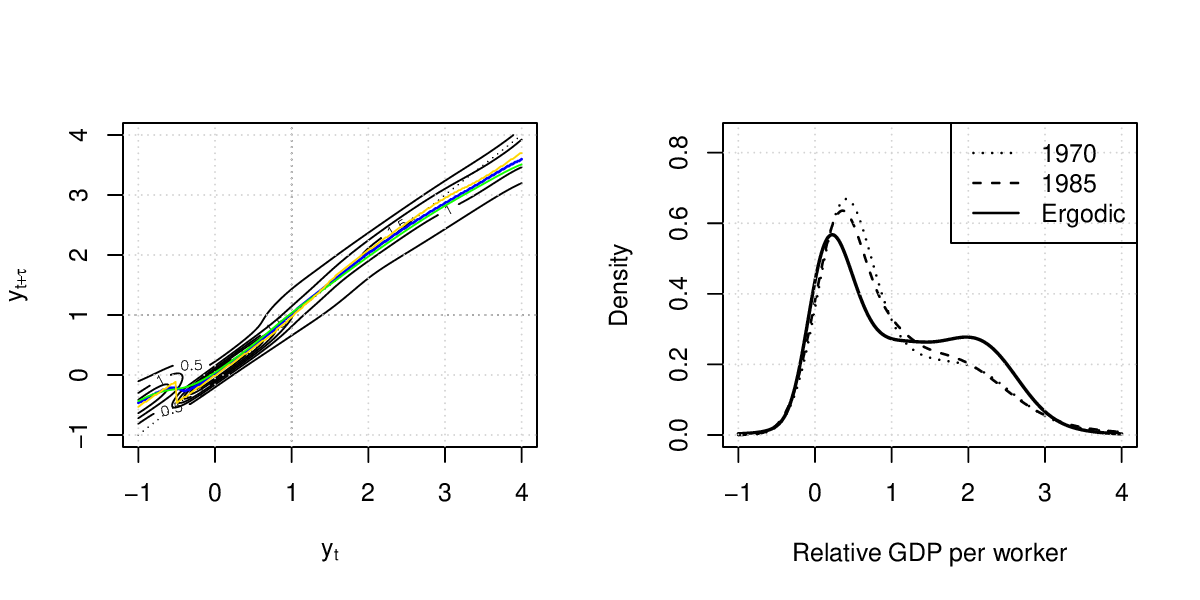}
		\caption{Period 1970-1985}
		\label{fig:stochastickernelergodicdistr19701985bandwidthoptimalalpha0}
	\end{subfigure}
	\begin{subfigure}{0.49\textwidth}
		\centering
		\includegraphics[width=\linewidth]{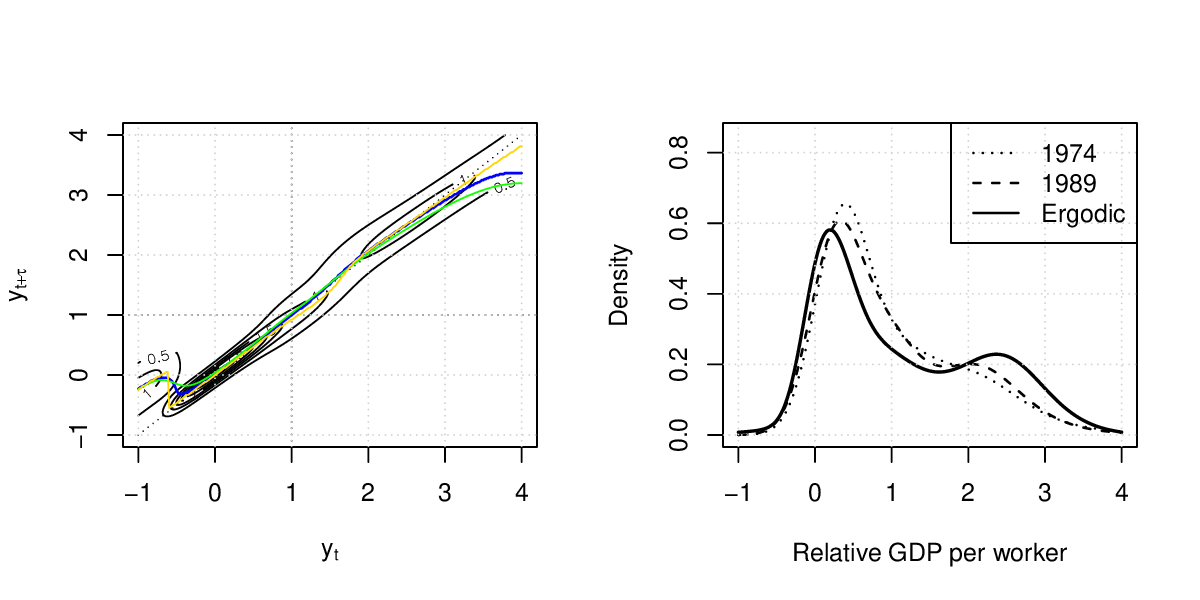}
		\caption{Period 1974-1989}
		\label{fig:stochastickernelergodicdistr19741989bandwidthoptimalalpha0}
	\end{subfigure}
	\begin{subfigure}{0.49\textwidth}
		\centering
		\includegraphics[width=\linewidth]{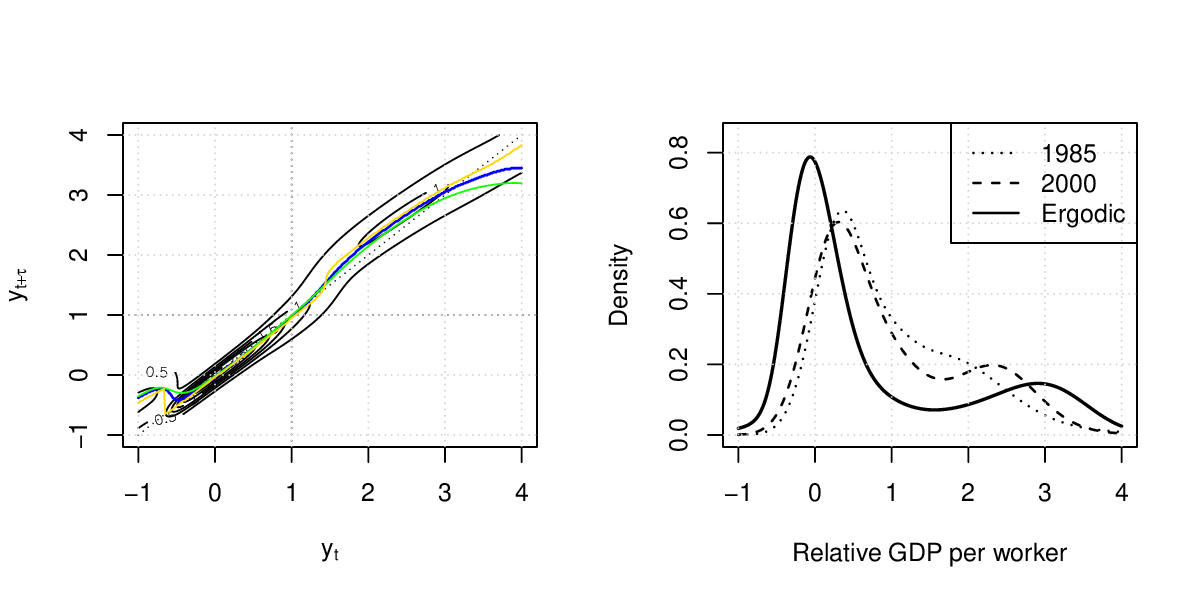}
		\caption{Period 1985-2000}
		\label{fig:stochastickernelergodicdistr19841999bandwidthoptimalalpha0}
	\end{subfigure}
	\begin{subfigure}{0.49\textwidth}
		\centering
		\includegraphics[width=\linewidth]{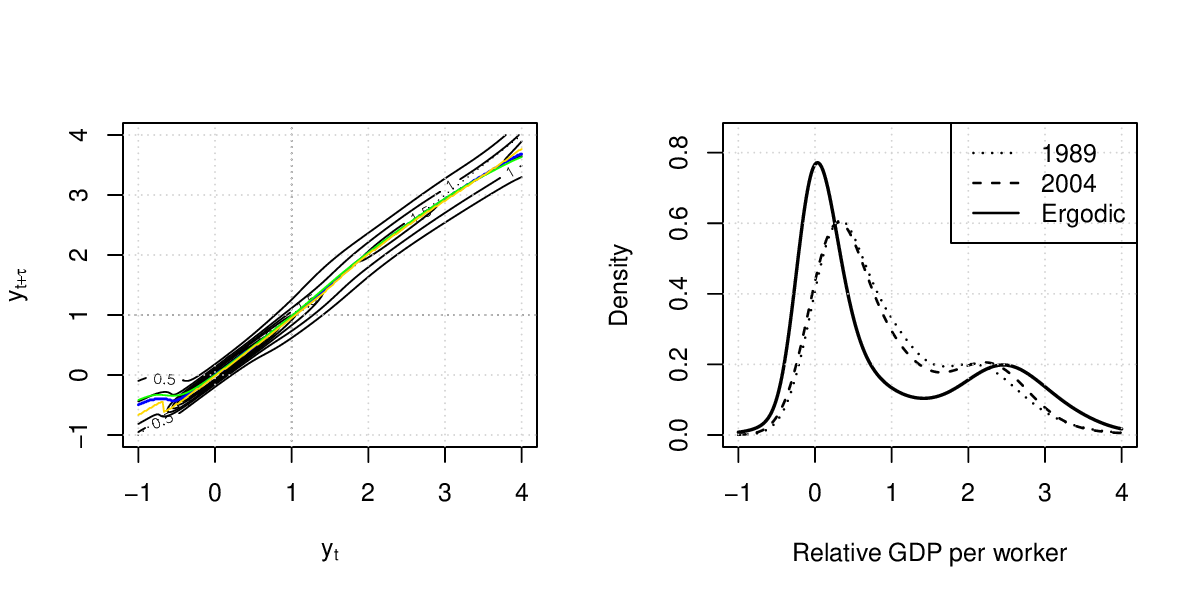}
		\caption{Period 1989-2004}
		\label{fig:stochastickernelergodicdistr19892004bandwidthoptimalalpha0}
	\end{subfigure}
	\begin{subfigure}{0.49\textwidth}
		\centering
		\includegraphics[width=\linewidth]{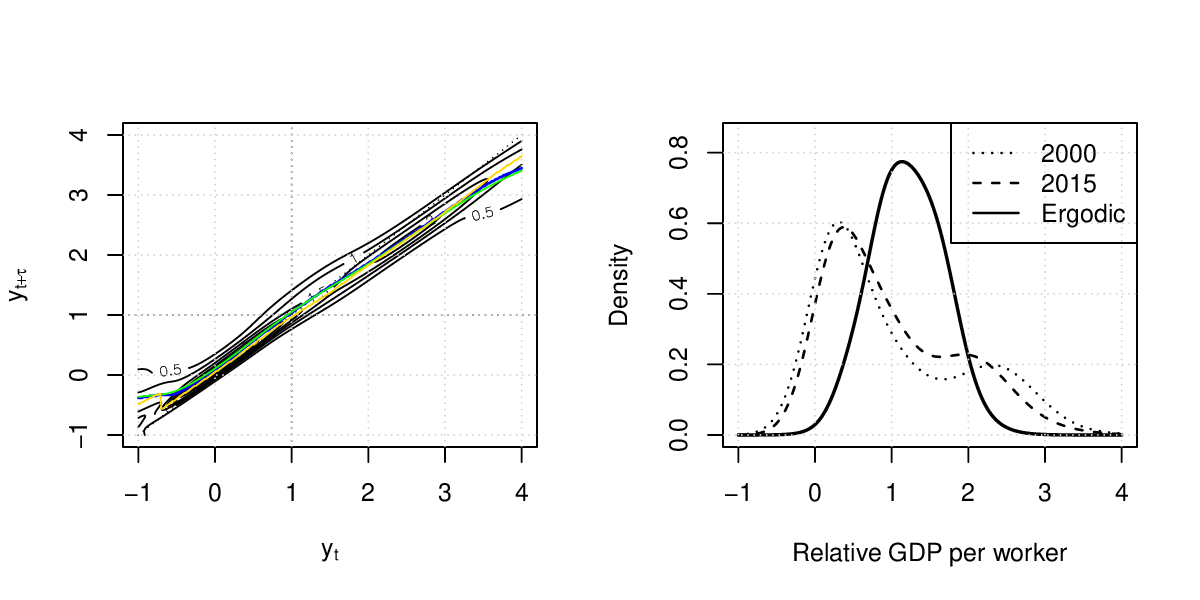}
		\caption{Period 2000-2015}
		\label{fig:stochastickernelergodicdistr20002015bandwidthoptimalalpha0}
	\end{subfigure}
	\begin{subfigure}{0.49\textwidth}
		\centering
		\includegraphics[width=\linewidth]{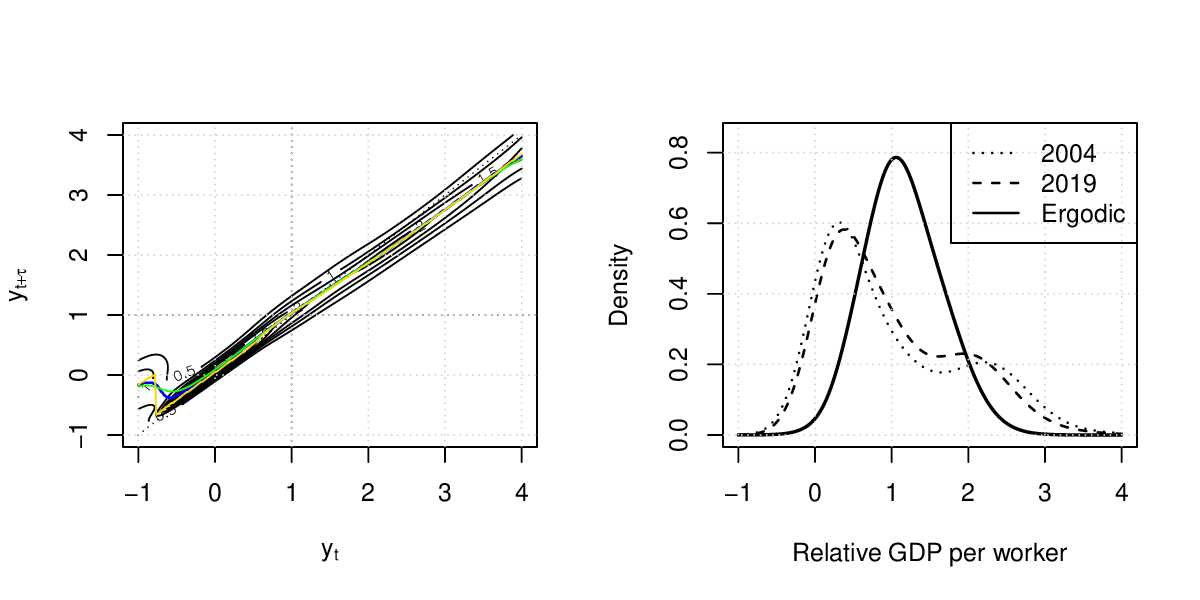}
		\caption{Period 2004-2019}
		\label{fig:stochastickernelergodicdistr20042019bandwidthoptimalalpha0}
	\end{subfigure}
	\caption{For each 15-year period, the first panel shows the estimated transition kernel of the cross country distribution of per capita GDP and the second shows the estimated density for the first and last year as well as the ergodic density implied by the estimated transition kernel. Estimation used a adaptive kernel estimator with a Gaussian kernel and the optimal Normal bandwidth \citep[p101]{silverman86}. The range of estimation is $[-1,4]$ in a regular grid of $100 \times 100$ evaluation points. The estimated transition kernel is calculated from the estimated joint distribution of the first and last years' per capita GDP levels as explained in the text. Authors' calculations using data from \citet{feenstra2015}.}
	\label{fig:15YearTransitions}
\end{figure}

\begin{table}[!htbp]
	\begin{tabular}{rHrr}
		\hline
		\hline
		\textbf{Test based on $L^1$}  & 1970-1985 & 1985-2000 & 2000-2015 \\
		\hline
		1970-1985 &  & 0.15 & \textbf{0.01} \\
		1985-2000 &  &  & \textbf{0.0}0 \\
		\hline
		\textbf{Test based on $H$} & 1970-1985 & 1985-2000 & 2000-2015 \\
		\hline
		1970-1985 &  & 0.23 & \textbf{0.02} \\
		1985-2000 &  &  & \textbf{0.00} \\
		\hline
		\\
		Number of obs & 102 & 102 & 102 \\
		\hline
		\hline
	\end{tabular}
	\begin{tabular}{rHrr}
		\hline
		\hline
		\textbf{Test based on $L^1$}  & 1974-1989 & 1989-2004 & 2004-2019 \\
		\hline
		1974-1989 &  & 0.06 & \textbf{0.00} \\
		1989-2004 &  &  & \textbf{0.00 }\\
		\hline
		\textbf{Test based on $H$}   & 1974-1989 & 1989-2004 & 2004-2019 \\
		\hline
		1974-1989 &  & 0.07 & \textbf{0.01} \\
		1989-2004 &  &  & \textbf{0.03} \\
		\hline
		\\
		Number of obs & 102 & 102 & 102 \\
		\hline
		\hline
	\end{tabular}
	\caption{ASL values computed for the null hypothesis of time-homogeneity for the indicated 15-year sub-periods using the procedure described in the text. Authors' calculations using data from \citet{feenstra2015}.}
	\label{tab:testTimeHomogeneity15YearSubperiods}
\end{table}

Table \ref{tab:testTimeHomogeneity15YearSubperiods} presents the ASL values for the time homogeneity tests that use the $L^1$ and $H$ distance measures supposing that the transition period is 15 years. We consider two sequences of 15-year transitions: 1970-1985, 1985-2000, 2000-2015; and 1974-1989, 1989-2004, 2004-2019. The comparisons of 1970-1985 and 1985-2000 and those of 1974-1989 and 1989-2004 do not yield rejections of the homogeneity hypothesis at the 5\% level, although the ASL values in the latter case imply rejections at the 10\% level. The comparisons of 1970-1985 with 2000-2015, of 1985-2000 with 2000-2015, of 1974-1989 with 2004-2019, and of 1989-2004 with 2004-2019 all result in rejections of homogeneity at the 5\% level. That is, similar to the 5-year and 10-year transition period results, there is evidence of homogeneity in the first half to two thirds of the sample which breaks down in the last half to one third. As Figure \ref{fig:15YearTransitions} shows, the lack of homogeneity found in the hypothesis tests is again evident in the ergodic distributions implied by the estimated transition kernels as those of the 2000-2015 and 2004-2019 periods are strongly single-peaked while those for the other 15-year transition periods in each sequence are bimodal.

\subsubsection{20-year sub-periods 	\label{sec:20YearSubPeriods}}

\begin{figure}[!htbp]
	\begin{subfigure}{0.49\textwidth}
	\centering
	\includegraphics[width=\linewidth]{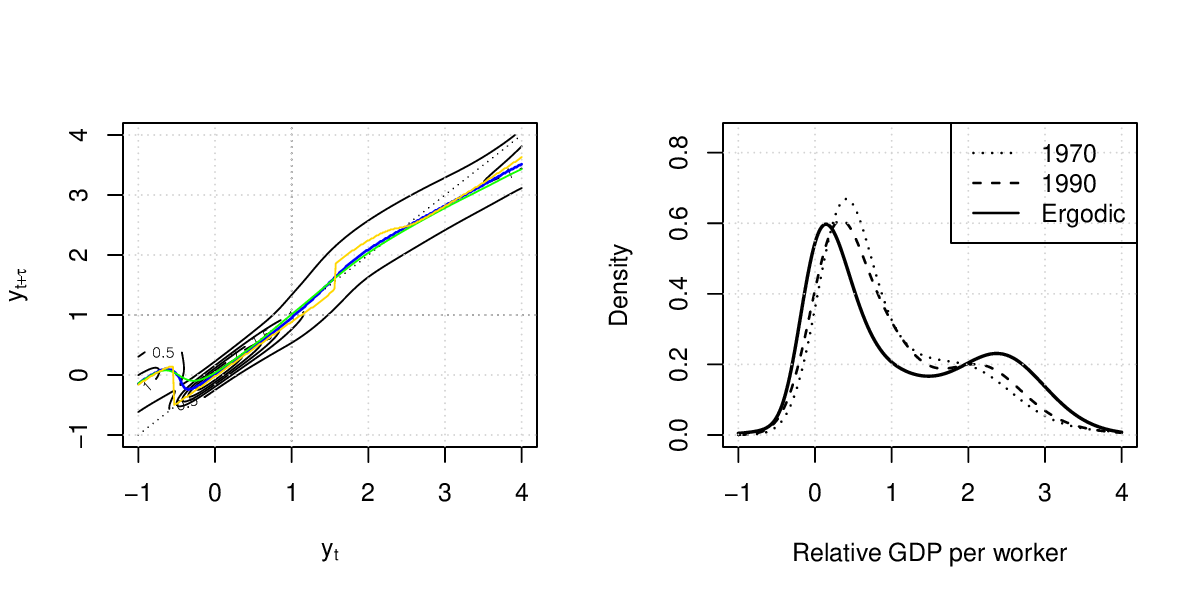}
	\caption{Period 1970-1990}
	\label{fig:stochastickernelergodicdistr19701989bandwidthoptimalalpha0}
\end{subfigure}
\begin{subfigure}{0.49\textwidth}
	\centering
	\includegraphics[width=\linewidth]{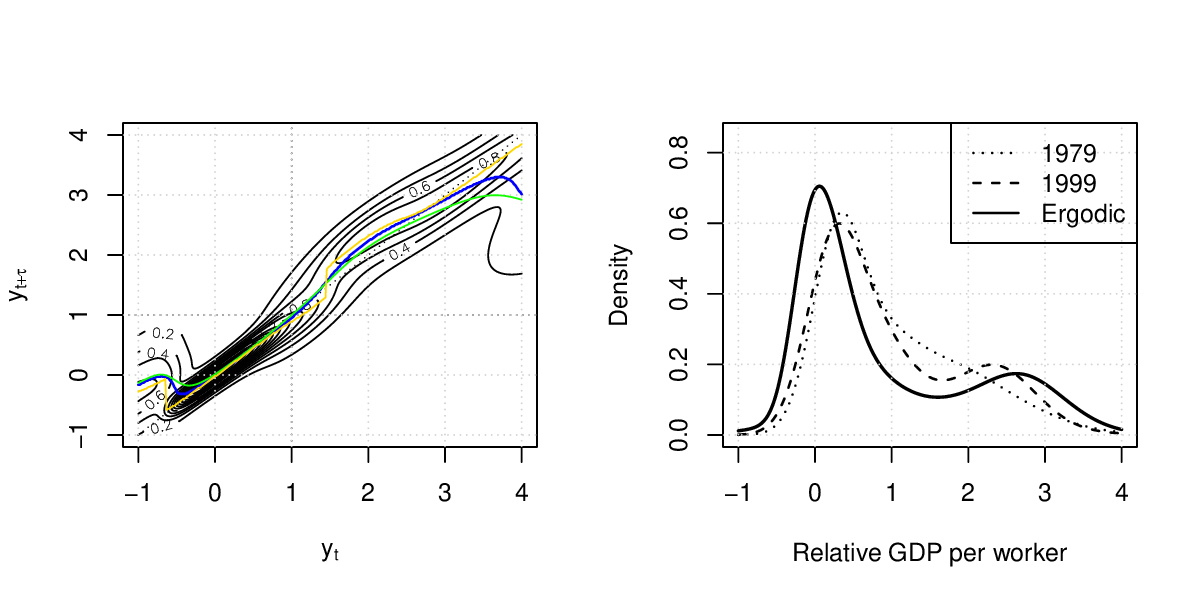}
	\caption{Period 1979-1999}
	\label{fig:stochasticKernelErgodicDistr_1979_1999bandwidth_optimalalpha_0.5}
\end{subfigure}
\begin{subfigure}{0.49\textwidth}
	\centering
	\includegraphics[width=\linewidth]{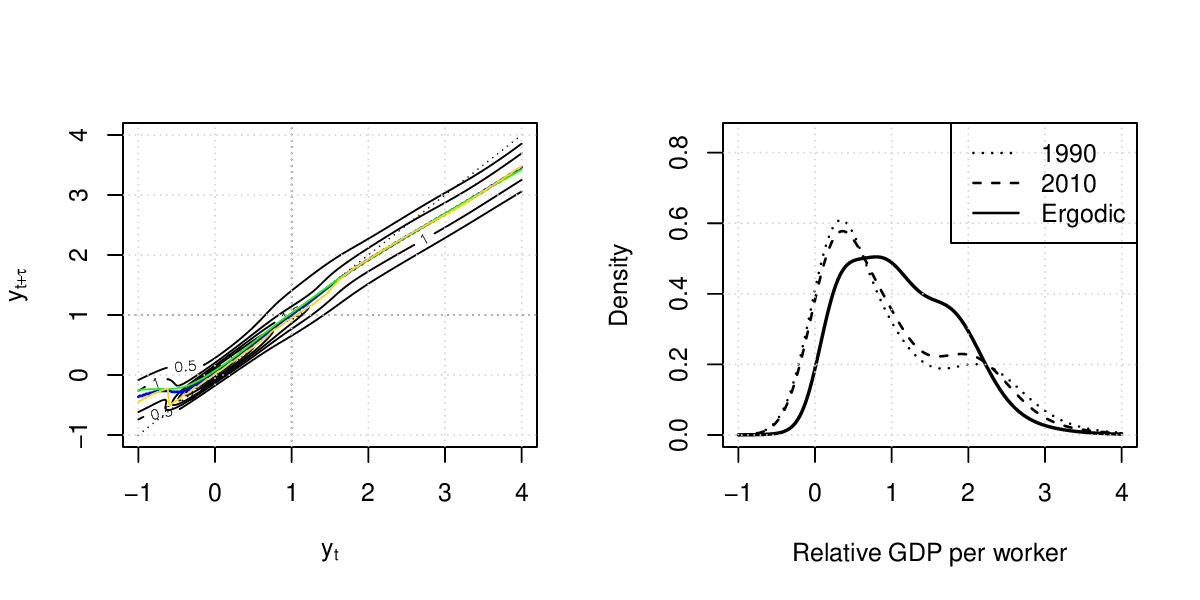}
	\caption{Period 1990-2010}
	\label{fig:stochastickernelergodicdistr19892009bandwidthoptimalalpha0}
\end{subfigure}
\begin{subfigure}{0.49\textwidth}
	\centering
	\includegraphics[width=\linewidth]{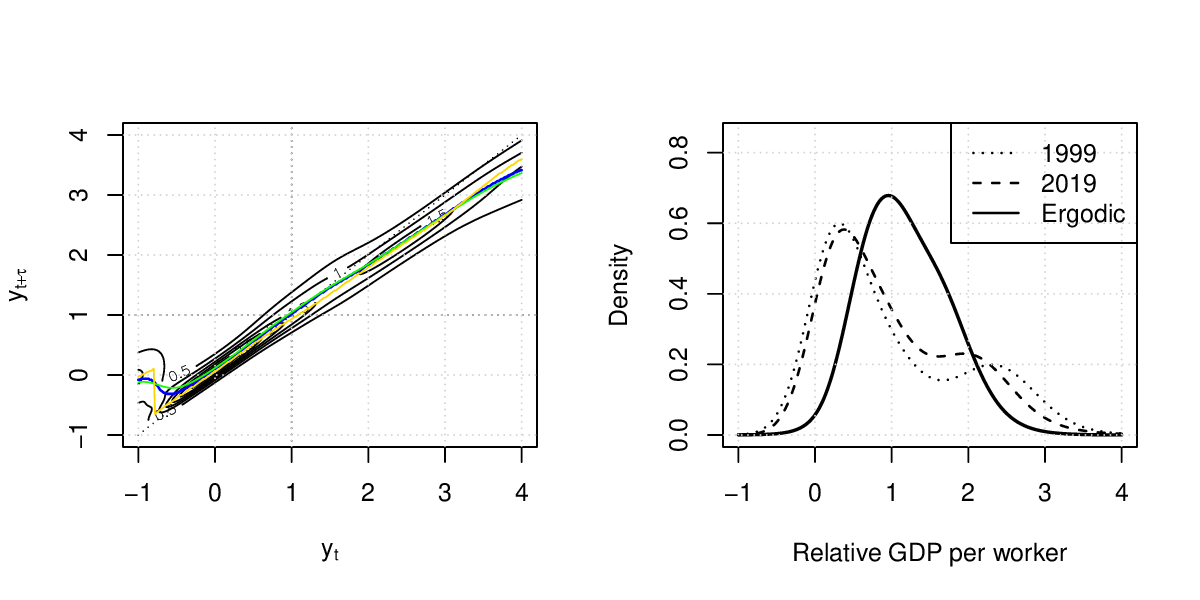}
	\caption{Period 1999-2019}
	\label{fig:stochasticKernelErgodicDistr_1999_2019bandwidth_optimalalpha_0.5}
\end{subfigure}
\caption{For each 20-year period, the first panel shows the estimated transition kernel of the cross country distribution of per capita GDP and the second shows the estimated density for the first and last year as well as the ergodic density implied by the estimated transition kernel. Estimation used a adaptive kernel estimator with a Gaussian kernel and the optimal Normal bandwidth \citet[p101]{silverman86}. The range of estimation is $[-1,4]$ in a regular grid of $100 \times 100$ evaluation points. The estimated transition kernel is calculated from the estimated joint distribution of the first and last years' per capita GDP levels as explained in the text. Authors' calculations using data from \citet{feenstra2015}.}
\label{fig:20YearTransitions}
\end{figure}

\begin{table}[!htbp]
	\centering
		\begin{tabular}{rHr}
		\hline
		\hline
		 \textbf{Test based on $L^1$} & 1970-1990 & 1990-2010 \\
		\hline
		1970-1990 &  & \textbf{0.00} \\
		\hline
		 \textbf{Test based on $H$}  & 1970-1990 & 1990-2010 \\
		\hline
		1970-1990 &  & \textbf{0.01} \\
		\hline
		\\
		Number of obs & 102 & 102\\
		\hline
		\hline
	\end{tabular}
	\begin{tabular}{rHr}
		\hline
		\hline
		 \textbf{Test based on $L^1$}    & 1979-1999 & 1999-2019 \\
		\hline
		1979-1999 &  & \textbf{0.00} \\
		\hline
	 \textbf{Test based on $H$}    & 1979-1999 & 1999-2019 \\
	\hline
	1979-1999 &  &\textbf{ 0.00} \\
	\hline
	\\
	Number of obs & 102 & 102\\
	   \hline
	      \hline
	\end{tabular}
	\caption{ASL values computed for the null hypothesis of time-homogeneity for the indicated 20-year sub-periods using the procedure described in the text. Authors' calculations using data from \citet{feenstra2015}.}
\label{tab:testTimeHomogeneity20YearSubperiods}
\end{table}

Table \ref{tab:testTimeHomogeneity20YearSubperiods} presents the ASL values for the time homogeneity tests that use the $L^1$ and $H$ distance measures supposing that the transition period is 20 years. We consider two sets of 20-year transitions: 1970-1990, 1990-2010; and 1979-1999, 1999-2019. All of the tests reject the null hypothesis at the 5\% level. The implied lack of homogeneity again evident in the ergodic distributions computed from the estimated transition kernels as those of the 1970-1990 and 1979-1999 periods are strongly twin-peaked while those for the 1990-2010 and 1999-2019 periods are unimodal, although that for 1990-2010 is not as strongly unimodal as that for 1999-2019, presumably because some pre-2000 data is used in its estimation.

\subsubsection{25-year sub-periods 	\label{sec:24YearSubPeriods}}

\begin{figure}[!htbp]
	\begin{subfigure}{0.49\textwidth}
		\centering
		\includegraphics[width=\linewidth]{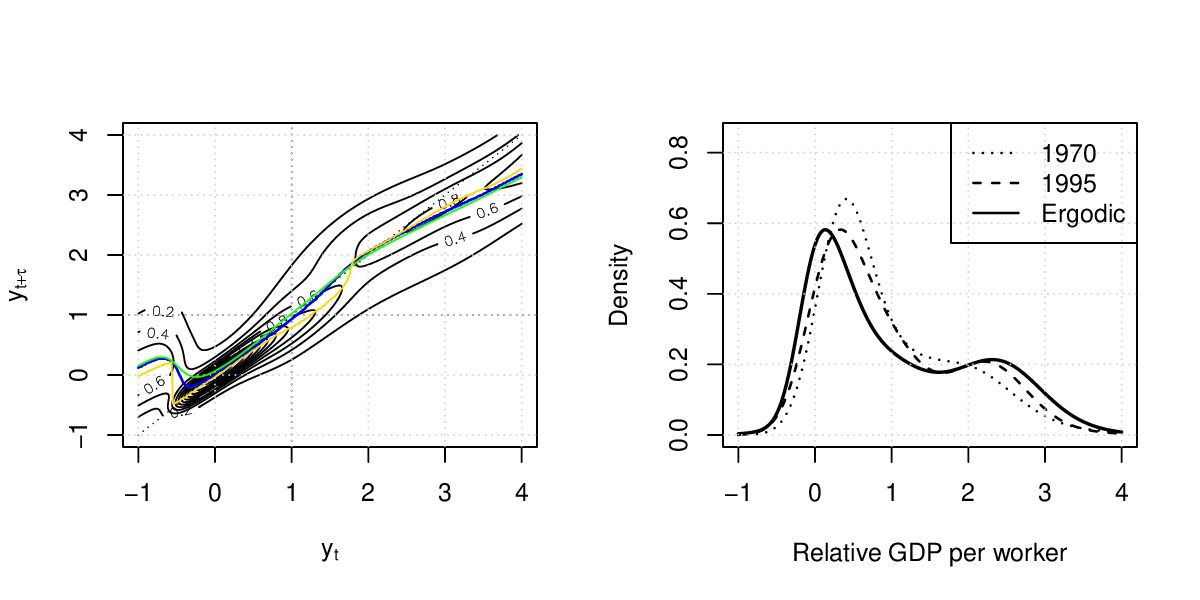}
		\caption{Period 1970-1995}
		\label{fig:stochasticKernelErgodicDistr_1970_1995bandwidth_optimalalpha_0.5}
	\end{subfigure}
	\begin{subfigure}{0.49\textwidth}
		\centering
		\includegraphics[width=\linewidth]{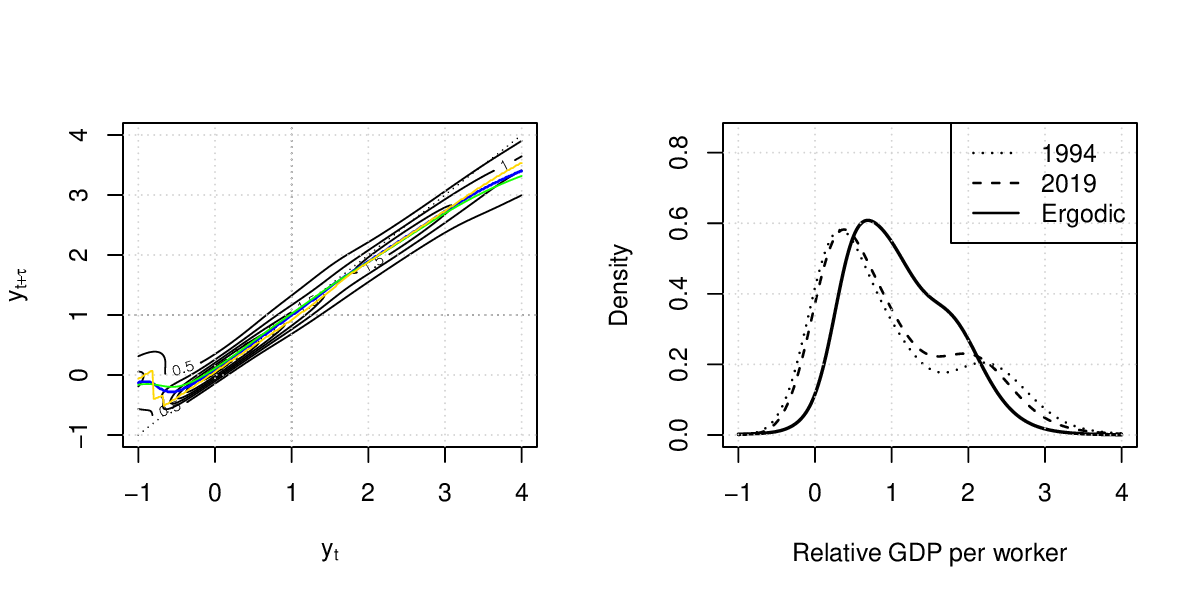}
		\caption{Period 1995-2019}
		\label{fig:stochasticKernelErgodicDistr_1994_2019bandwidth_optimalalpha_0.5}
	\end{subfigure}
\caption{For each 25-year period, the first panel shows the estimated transition kernel of the cross country distribution of per capita GDP and the second shows the estimated density for the first and last year as well as the ergodic density implied by the estimated transition kernel. Estimation used a adaptive kernel estimator with a Gaussian kernel and the optimal Normal bandwidth \citep[p101]{silverman86}. The range of estimation is $[-1,4]$ in a regular grid of $100 \times 100$ evaluation points. The estimated transition kernel is calculated from the estimated joint distribution of the first and last years' per capita GDP levels as explained in the text. Authors' calculations using data from \citet{feenstra2015}.}
\label{fig:25YearTransitions}
\end{figure}

\begin{table}[!htbp]
	\centering
	\begin{tabular}{rHr}
		\hline
		\hline
		 \textbf{Test based on $L^1$}    & 1970-1995 & 1994-2019 \\
		\hline
		1970-1995 &  & \textbf{0.00 }\\
		\hline
		  \\
		\textbf{Test based on $H$}  & 1970-1995 & 1994-2019 \\
		\hline
		1970-1995 &  & \textbf{0.00} \\
		\hline
		\\
		Number of obs & 102 & 102\\
	 \hline
	\hline
	\end{tabular}
	\caption{ASL values computed for the null hypothesis of time-homogeneity for the indicated 25-year sub-periods using the procedure described in the text. Authors' calculations using data from \citet{feenstra2015}.}
	\label{tab:testTimeHomogeneity24YearSubperiods}
\end{table}

The tests that use the $L^1$ and $H$ distance measures supposing that the transition period is 25 years presented in Table \ref{tab:testTimeHomogeneity24YearSubperiods}, which compare the dynamics from 1970-1995 with those from 1994-2019, result in rejections of time homogeneity at the 5\% level. Figure \ref{fig:25YearTransitions} shows that, as with the shorter transition lengths, that while the ergodic distributions implied by the estimated transition kernel for earlier period (1970-1995) is bimodal, that for the latter period (1994-2019) is unimodal,  although arguably not as strongly so as are those from post-2000 data for other transition lengths, presumably because some pre-2000 data is involved in the estimation.

\subsubsection{Summary of homogeneity test results	\label{sec:Homogeneitysummary}}

In this subsection we have tested the time homogeneity of the process governing the evolution of the cross-country distribution of per capita income for transition lengths of 5, 10, 15, 20 and 25 years using data from the 1970-2019 period. For the transition lengths of 5, 10, and 15 years, we find evidence that the process was time homogeneous during the first part of the sample, that is from 1970 to 1995 or 2000. During this period, the dynamics of the process imply a long-run distribution of per capita income that is twin-peaked, similar to the distributions estimated for individual years. We find strong evidence of a large shift in the process in the 1995-2000 period which ushers in a decade or so of dynamics broadly consistent with a single-peaked long-run distribution.  The 5-year transition length estimates suggest an apparent return to dynamics similar to those in the first part of the sample at some point in the 2010s.  The tests for the 20-year and 25 year transition lengths are consistent with the finding of a shift in the process in the 1995-2000 period from one with dynamics implying a bimodal long-run distribution to one with dynamics implying a unimodal long-run distribution. We take a more detailed look at the distribution dynamics in different sub-periods after testing the order of the process in the next section.

\subsection{Testing the order of process \label{sec:OrderProcess}}

The test results in Section \ref{sec:timeHomogeneity} suggest the presence of three sub-periods where the process is time homogeneous, i.e. 1970-1995, 2000-2010, and 2009-2019, although the difference between 2000-2010 and 2009-1019 is a good deal less stark than that between 1970-1995 and 2000-2010. The lengths of these subperiods mean that we can test the first-orderedness of the process (conditional on homogeneity) for 5-year and 10-year transitions in the first sub-period but only for 5-year transitions in the last two sub-periods. We exploit the apparent homogeneity in the 1970-1995 period to increase the sample sizes for both the 5-year and 10-year transition period tests by, in the case of the 5-year tests, taking as the transition pairs as 1970-1975 and 1975-1980, 1971-1976 and 1976-1981, ..., 1985-1990 and 1990-1995 for a total of 1632 observations. For each of the 2000-2010 and 2009-1019 samples there are 102 observations. In the case of the 10-year tests we take the transition pairs as 1970-1980 and 1980-1990, 1971-1981 and 1981-1991, ..., 1975-1985 and 1985-1995 for a total of 612 observations.

Table \ref{tab:testPWTFirstOrderProcess} presents the results. In the 1970-1995 sub-period, for the 5-year transitions, the hypothesis $H_0$ that the process is first order is rejected at the 5\% significance level using the $L^2$ and $L^\infty$ distance measures and at the 10\% level using the $H$ distance measure. For the 10-year transitions, the only rejection that occurs is at the 10\% level using the $L^\infty$ distance measure. These results are consistent with the process describing the evolution of the cross-country distribution of per capita income being an homogenous, first-order process during the 1970-1995 period provided the transition length is taken to be 10 years.
Using the 5-year transition length, for the 2000-2010 period, we cannot reject the hypothesis $H_0$ that the process is first-order but, for the 2009-2019, that hypothesis is rejected at the 10\% level using the $H$ distance measure. While these results, conditional on the homogeneity of the process in these periods, are consistent with a first-order process, some caution must be using in reaching that conclusion given the relatively low power of this tests in small samples and the weakness of the evidence for the homogeneity of the process after 2000.

\begin{table}[!htbp]
		\centering
	\begin{tabular}{ccHcc}
		\hline
		\hline
		& 1970-1995 & 1995-1970  & 2000-2010 & 2009-2019  \\
		\hline
		\textbf{5-year transitions} & & \\
		$L^1$ & 0.09 & 0.35 & 0.37& 0.14\\
		$L^2$ & 0.04 & 0.04  & 0.29 & 0.33 \\
		$L^\infty$ & 0.00 & 0.00 &   0.24 & 0.58 \\
		$H$ & 0.27 & 0.59 &  0.62& 0.06 \\
		\hline
		Number of obs & 1632 & 1632 & 102 & 102 \\
		\\
		\hline
		\textbf{10-year transitions} & & \\
		$L^1$ & 0.32 & 0.04\\
		$L^2$ & 0.21  & 0.01\\
		$L^\infty$ & 0.06 & 0.01\\
		$H$ & 0.42 & 0.11 \\
		\hline
		Number of obs & 612 & 612\\
		\hline
		\hline
	\end{tabular}
	\caption{The ASL based on 1000 bootstraps of the null hypothesis of first-order process for the three subperiods 1970-1995, 2000-2010, and 2009-2019. The density estimation is made by an adaptive Normal kernel using the rule of the optimal Normal bandwidth \citep[101]{silverman86}. The range of estimation is $[-1,4]$ in a regular grid of $100 \times 100$ evaluation points.}
	\label{tab:testPWTFirstOrderProcess}
\end{table}

\subsection{The distribution dynamics in the sub-periods \label{sec:timeHomogeneousFirstOrderMarkovProcessPWT}}

The main findings of Sections \ref{sec:timeHomogeneity} and \ref{sec:OrderProcess} suggest the existence of three sub-periods where distribution dynamics appears to follow a first-order time-homogeneous Markov process, i.e. 1970-1995, 2000-2010 and 2009-2019. For the first of these periods, the hypothesis seems to hold for the 10-year transition length but not for the 5-year transition length while for the other two, it seems to hold for the 5-year transition length. We are, however, far more confident about the conclusion for the 1970-1995 period given the far larger sample size available for that period and the possible lower power of our proposed test of the order of the process in small samples. In this section we investigate the distribution dynamics for each of the sub-periods assuming that the process is first-order and time-homogeneous within each sub-period.

For the 1970-1995, 2000-2010, and 2009-2019 sub-periods we consider transition lengths of 5 and 10 years, those for which we've found evidence of within-period homogeneity and first-orderedness. To increase the sample size, we use overlapping transitions so, for the 5-year transition length data in the 1970-1995 sub-period, for example, we pool the 1970-1975, 1971-1976, 1972-1977, $\ldots$, 1990-1995 transition data to give a sample of 2142 observations. For the 10-year transition length, this procedure yields 1122 observations. Similarly, for the 2000-2010, and 2009-2019 sub-periods we have 612 and 102 observations for the 5-year and 10-year transition lengths respectively. For each transition length in each sub-period, we compute the transition kernel and the implied ergodic density using the a sample of data constructed in this way. Rather than present the estimated transition kernels, we use the ergodic distribution as a summary statistic for the transition dynamics. We bootstrap the 90\% confidence intervals for the ergodic distributions by sampling with replacement from the appropriate pooled dataset and repeating the estimation of the ergodic distribution. Having done this 1000 times, the upper and lower limits of the intervals are, at each point on the horizontal axis, the 50$^{\text{th}}$ largest and the 50$^{\text{th}}$ smallest values of the estimated ergodic distribution.

\begin{figure}[!htbp]
	\centering
	\begin{subfigure}{0.32\linewidth}
	\includegraphics[width=\linewidth]{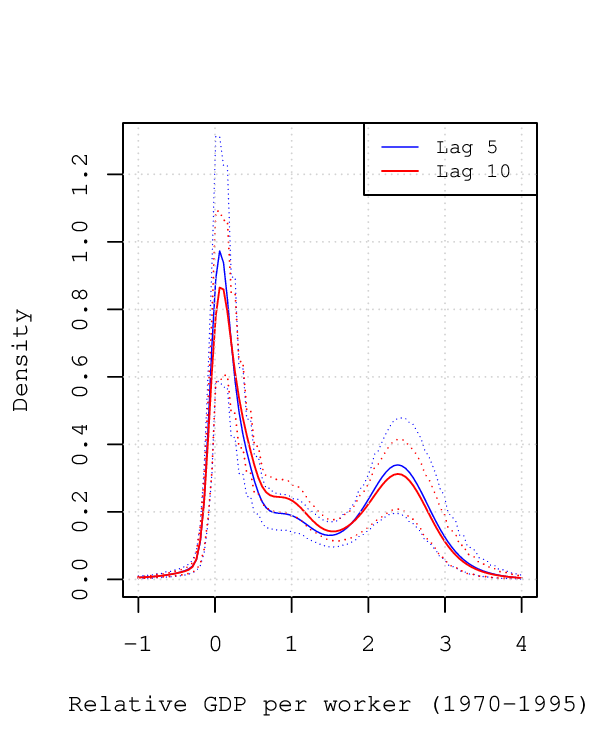}
	\caption{Sub-period 1970-1995}
	\label{fig:comparisonergodicdistributionsdifferentlagtwomainperiods19701995}
	\end{subfigure}
\begin{subfigure}{0.33\linewidth}
	\centering
	\includegraphics[width=\linewidth]{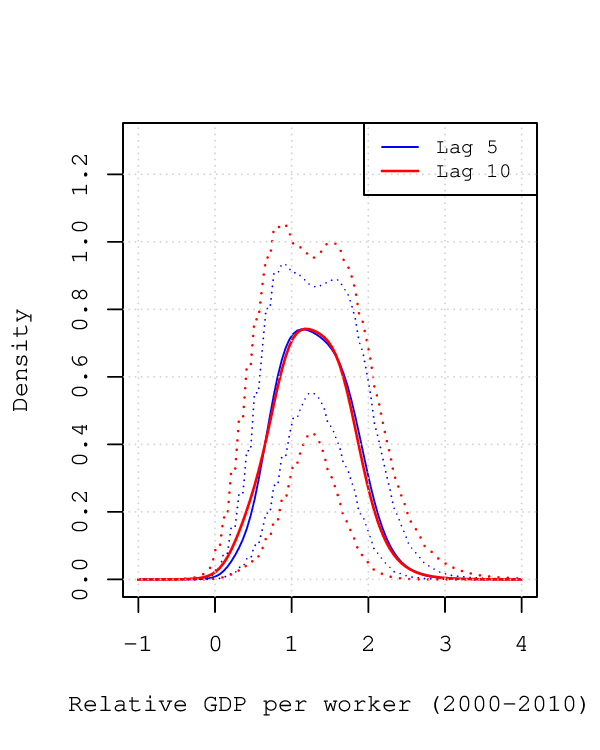}
	\caption{Sub-period 2000-2010}
	\label{fig:comparisonergodicdistributionsdifferentlagtwomainperiods20002010}
\end{subfigure}
\begin{subfigure}{0.33\linewidth}
	\centering
	\includegraphics[width=\linewidth]{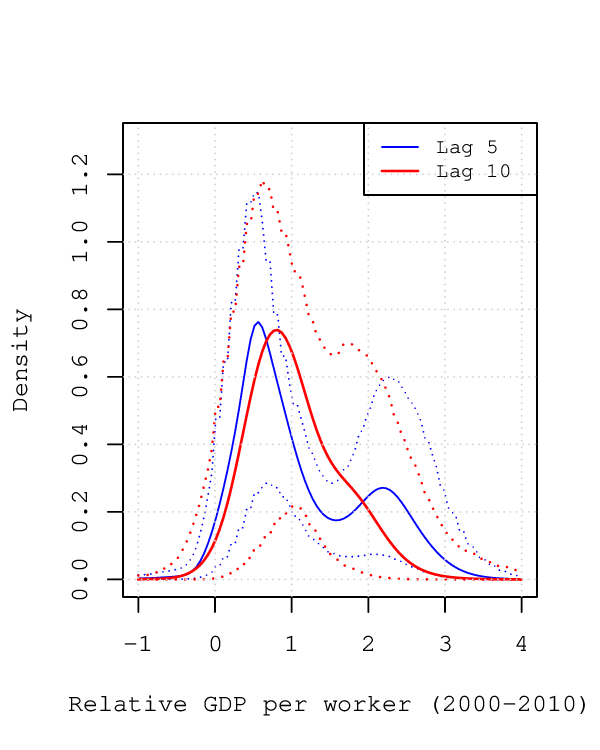}
	\caption{Sub-period 2009-2019}
	\label{fig:comparisonergodicdistributionsdifferentlagtwomainperiods20092019}
\end{subfigure}
\caption{The estimate of ergodic distributions and their 90\% confidence bands for different transition lengths and sub-periods.}
\label{fig:ergodicDistrSubperiods}
\end{figure}

Figure \ref{fig:comparisonergodicdistributionsdifferentlagtwomainperiods19701995} shows the estimated ergodic distributions for both transition lengths and the bootstrapped 90\% confidence bands for the 1970-1995 sub-period. For this period, the estimated ergodic distributions are very similar being bimodal with both modes, and the antimode between them, occurring at very close to the same value of relative GDP per worker. The estimated confidence intervals imply that the estimated bimodality is statistically significant in both cases. This is consistent with the earlier estimated ergodic densities using various sub-periods of the pre-2000 data for various transition lengths and is suggestive of the existence of convergence clubs, although that conclusion also requires consideration of the extent of the mobility within the distribution as explained in \citet{pittau2010mixture}.

For both of the transition lengths, the estimated ergodic densities for the 2000-2010 sub-period presented in Figure  \ref{fig:comparisonergodicdistributionsdifferentlagtwomainperiods20002010} show clear unimodality, which, despite their width, the confidence intervals suggest is statistically significant. This unimodality is suggestive of a single convergence club and thus absolute convergence consistent with the short-term $\beta$-convergence found by \citet{patel2021new} and \citet{kremer2022converging} for the same period.\footnote{See Figures 1 and 2 in \citet{kremer2022converging}. \citet{imam2025} also find an inhomogeneity in cross-country income dynamics in the mid-1990s that suggests a change from a non-convergent to a possibly convergent process. They attribute this change to reductions in the likelihood and duration of economic crises and note the implied fragility of the tendency towards convergence due to its dependence on the behaviour of the crisis process.} The unimodality shown in Figure \ref{fig:comparisonergodicdistributionsdifferentlagtwomainperiods20002010}reflects that found in Figures \ref{fig:5YearTransitions} to \ref{fig:20YearTransitions} which all show unimodal ergodic distributions for transition kernels estimated using data from around 2000-2010. Figure \ref{fig:stochastickernelergodicdistr20002010bandwidthoptimalalpha0} shows the estimated cross-country distributions of per capita GDP in 2000 and 2010. Both are bimodal but the right-hand mode for the 2010 distribution lies slightly to the left of that for 2000, presumably reflecting the tendency to convergence evident in this decade, while the left-hand modes for both distributions are almost coincident. Together these observations suggest that the dynamics that imply a unimodal ergodic distribution take quite a while to affect the shape of the observed cross-country distribution of per capita GDP and whatever effect did occur mattered more for the rich countries than for the poor countries.

In the 2009-2019 sub-period, signs of bimodality, suggestive of divergence, re-emerge. The estimated ergodic distribution for the 10-year transition length shown in Figure \ref{fig:comparisonergodicdistributionsdifferentlagtwomainperiods20092019} is unimodal but that for the 5-year transition length is bimodal and similar to those estimated for the 1970-1995 sub-period. The estimated confidence interval for the 5-year transition length ergodic density is, however, too wide to imply that the bimodality is statistically significant. Nonetheless, the estimates for the 2009-2019 period suggest that the tendency to convergence observed in the 2000s may have stopped or at least weakened around 2010 and the estimated cross-country distributions of per capita GDP in 2009 and 2019 shown in Figure \ref{fig:stochastickernelergodicdistr20092019bandwidthoptimalalpha0} are almost coincident. While \cite{kremer2022converging} consider data only through 2007, Figure 2 of \cite{patel2021new} contains evidence of this reversal in the dynamics.

\subsection{The distribution dynamics conditional on initial per capita GDP \label{sec:conditionaloninitalpercapitaGDP}}

Our discussion of Figure \ref{fig:normalizedQuintileBoundaries} and Figure \ref{fig:estimatedCrossSectionalDistributionsDifferentYears} in Section \ref{sec:datamotivation} hinted at stability in the evolution of the cross-country distribution of per capita GDP early in our sample which is followed by instability especially among the higher income countries as both the estimated quintile boundaries and the estimated cross-sectional densities exhibit much more stability in the low income parts of the distribution than in the high income parts. This observation and the results presented above, which confirm the instability in the evolution of the distribution post-1995, invite investigation of the stability conditional on initial per capita GDP values. To do this, we divide our 102 country sample into three sub-samples of 34 countries according to relative per capita GDP levels in 1970 with the 34 countries with the lowest incomes in the "low income" sub-sample, the 34 countries with the highest incomes in the "high-income" sub-sample and the 34 countries in the middle of the distribution in the "medium-income" sub-sample.

For each of the three sub-samples, we repeat the homogeneity tests for the 5-year and 10-year transitions, that is, those for which we've documented the apparent shift from a divergent process in the mid-1990s to a seemingly convergent process until about 2010 before a return to a possibly divergent process. Tables\ref{tab:testTimeHomogeneity5yearslowmediumhigh} and \ref{tab:testTimeHomogeneity10yearslowmediumhigh} present the results. We report only the results using the $L^1$ and $H$ distance measures as they seem to have the best size and power properties of the four distance measures that we employ and we are mindful of the small samples here due to the dividing of the sample.

\begin{table}[htbp]
	\centering
	{\tiny
		\begin{tabular}{rHrrrrrrrrr}
			\hline
			\textbf{$L^1$ for low-income countries} & 1970-1975 & 1975-1980 & 1980-1985 & 1985-1990 & 1990-1995 & 1995-2000 & 2000-2005 & 2005-2010 & 2010-2015 & 2014-2019 \\
			\hline
			1970-1975 &  & 0.67 & 0.34 & 0.54 & 0.70 & 0.60 & 0.46 & 0.13 & 0.06 & 0.26 \\
			1975-1980 &  &  & 0.39 & 0.67 & 0.89 & 0.34 & 0.42 & 0.15 & \textbf{0.04} & 0.41 \\
			1980-1985 &  &  &  & 0.32 & 0.44 & 0.35 & 0.98 & 0.57 & 0.63 & 0.53 \\
			1985-1990 &  &  &  &  & 0.78 & 0.24 & 0.32 & 0.25 & \textbf{0.03} & 0.26 \\
			1990-1995 &  &  &  &  &  & 0.33 & 0.32 & 0.12 & \textbf{0.04} & 0.44 \\
			1995-2000 &  &  &  &  &  &  & 0.37 & 0.06 & 0.07 & 0.23 \\
			2000-2005 &  &  &  &  &  &  &  & 0.49 & 0.82 & 0.43 \\
			2005-2010 &  &  &  &  &  &  &  &  & 0.23 & 0.46 \\
			2010-2015 &  &  &  &  &  &  &  &  &  & 0.19 \\
			\hline
		\end{tabular}
		\begin{tabular}{rHrrrrrrrrr}
			\hline
			\textbf{$L^1$ for medium-income income} & 1970-1975 & 1975-1980 & 1980-1985 & 1985-1990 & 1990-1995 & 1995-2000 & 2000-2005 & 2005-2010 & 2010-2015 & 2014-2019 \\
			\hline
			1970-1975 &  & 0.71 & 0.69 & 0.70 & 0.68 & 0.09 & 0.62 & 0.15 & \textbf{0.01} & \textbf{0.02} \\
			1975-1980 &  &  & 0.22 & 0.31 & 0.32 & 0.03 & 0.37 & \textbf{0.04} & \textbf{0.00} & \textbf{0.00} \\
			1980-1985 &  &  &  & 0.50 & 0.44 & 0.32 & 0.91 & 0.37 & 0.12 & 0.09 \\
			1985-1990 &  &  &  &  & 0.66 & 0.20 & 0.83 & \textbf{0.04} & \textbf{0.01} & \textbf{0.04} \\
			1990-1995 &  &  &  &  &  & 0.16 & 0.58 & \textbf{0.03} & \textbf{0.01} & \textbf{0.03} \\
			1995-2000 &  &  &  &  &  &  & 0.35 & \textbf{0.01} & \textbf{0.02} & 0.23 \\
			2000-2005 &  &  &  &  &  &  &  & 0.12 & 0.07 & 0.14 \\
			2005-2010 &  &  &  &  &  &  &  &  & 0.61 & 0.06 \\
			2010-2015 &  &  &  &  &  &  &  &  &  & 0.35 \\
			\hline
		\end{tabular}
		\begin{tabular}{rHrrrrrrrrr}
			\hline
			\textbf{$L^1$ for high-income countries} & 1970-1975 & 1975-1980 & 1980-1985 & 1985-1990 & 1990-1995 & 1995-2000 & 2000-2005 & 2005-2010 & 2010-2015 & 2014-2019 \\
			\hline
			1970-1975 &  & 0.98 & 0.81 & 0.19 & 0.68 & \textbf{0.01} & 0.12 & 0.12 & \textbf{0.01} & \textbf{0.00} \\
			1975-1980 &  &  & 0.86 & 0.32 & 0.56 & \textbf{0.01} & 0.19 & 0.12 & 0.07 & \textbf{0.02} \\
			1980-1985 &  &  &  & 0.47 & 0.75 & \textbf{0.03} & 0.40 & 0.17 & 0.15 & \textbf{0.04} \\
			1985-1990 &  &  &  &  & 0.20 & \textbf{0.02} & \textbf{0.03} & \textbf{0.01} & \textbf{0.02} & \textbf{0.01} \\
			1990-1995 &  &  &  &  &  & 0.08 & 0.12 & 0.05 & 0.31 & 0.06 \\
			1995-2000 &  &  &  &  &  &  & \textbf{0.00} & \textbf{0.00} & 0.41 & 0.27 \\
			2000-2005 &  &  &  &  &  &  &  & 0.54 &\textbf{ 0.02} & \textbf{0.00 }\\
			2005-2010 &  &  &  &  &  &  &  &  &\textbf{ 0.01} & \textbf{0.00} \\
			2010-2015 &  &  &  &  &  &  &  &  &  & 0.90 \\
			\hline
		\end{tabular}
		\begin{tabular}{rHrrrrrrrrr}
			\hline
			\textbf{$H$ for low-income countries} & 1970-1975 & 1975-1980 & 1980-1985 & 1985-1990 & 1990-1995 & 1995-2000 & 2000-2005 & 2005-2010 & 2010-2015 & 2014-2019 \\
			\hline
			1970-1975 &  & 0.58 & 0.45 & 0.43 & 0.63 & 0.53 & 0.48 & 0.27 & 0.07 & 0.20 \\
			1975-1980 &  &  & 0.47 & 0.38 & 0.85 & 0.43 & 0.52 & 0.19 & 0.07 & 0.32 \\
			1980-1985 &  &  &  & 0.38 & 0.58 & 0.45 & 0.93 & 0.53 & 0.66 & 0.34 \\
			1985-1990 &  &  &  &  & 0.66 & 0.40 & 0.51 & 0.36 & 0.06 & 0.32 \\
			1990-1995 &  &  &  &  &  & 0.52 & 0.49 & 0.13 & 0.06 & 0.23 \\
			1995-2000 &  &  &  &  &  &  & 0.42 & 0.08 & 0.08 & 0.27 \\
			2000-2005 &  &  &  &  &  &  &  & 0.51 & 0.69 & 0.36 \\
			2005-2010 &  &  &  &  &  &  &  &  & 0.20 & 0.49 \\
			2010-2015 &  &  &  &  &  &  &  &  &  & 0.16 \\
			\hline
		\end{tabular}
		\begin{tabular}{rHrrrrrrrrr}
			\hline
			\textbf{$H$ for medium-income countries} & 1970-1975 & 1975-1980 & 1980-1985 & 1985-1990 & 1990-1995 & 1995-2000 & 2000-2005 & 2005-2010 & 2010-2015 & 2014-2019 \\
			\hline
			1970-1975 &  & 0.61 & 0.77 & 0.61 & 0.73 & 0.14 & 0.63 & 0.25 & 0.06 & 0.06 \\
			1975-1980 &  &  & 0.19 & 0.24 & 0.46 & 0.03 & 0.53 & 0.09 & \textbf{0.03} & \textbf{0.00} \\
			1980-1985 &  &  &  & 0.51 & 0.36 & 0.36 & 0.67 & 0.30 & 0.27 & 0.17 \\
			1985-1990 &  &  &  &  & 0.46 & 0.24 & 0.60 & \textbf{0.04} & \textbf{0.03} & 0.07 \\
			1990-1995 &  &  &  &  &  & 0.17 & 0.59 & \textbf{0.03 }& 0.06 & 0.06 \\
			1995-2000 &  &  &  &  &  &  & 0.27 & \textbf{0.01} & 0.07 & 0.37 \\
			2000-2005 &  &  &  &  &  &  &  & 0.12 & 0.22 & 0.12 \\
			2005-2010 &  &  &  &  &  &  &  &  & 0.68 & \textbf{0.04} \\
			2010-2015 &  &  &  &  &  &  &  &  &  & 0.31 \\
			\hline
		\end{tabular}
		\begin{tabular}{rHrrrrrrrrr}
			\hline
			\textbf{$H$ for high-income countries} & 1970-1975 & 1975-1980 & 1980-1985 & 1985-1990 & 1990-1995 & 1995-2000 & 2000-2005 & 2005-2010 & 2010-2015 & 2014-2019 \\
			\hline
			1970-1975 &  & 0.93 & 0.51 & 0.16 & 0.36 & \textbf{0.01} & 0.22 & 0.07 & \textbf{0.02} & \textbf{0.01} \\
			1975-1980 &  &  & 0.76 & 0.49 & 0.33 & 0.02 & 0.32 & 0.11 & 0.08 & \textbf{0.04} \\
			1980-1985 &  &  &  & 0.50 & 0.57 & 0.13 & 0.56 & 0.23 & 0.26 & 0.12 \\
			1985-1990 &  &  &  &  & 0.23 & 0.07 & 0.09 & \textbf{0.04} & 0.06 & \textbf{0.03} \\
			1990-1995 &  &  &  &  &  & 0.06 & 0.22 & 0.08 & 0.41 & 0.15 \\
			1995-2000 &  &  &  &  &  &  & \textbf{0.00} & \textbf{0.00} & 0.54 & 0.41 \\
			2000-2005 &  &  &  &  &  &  &  & 0.56 & \textbf{0.04 }&\textbf{ 0.01} \\
			2005-2010 &  &  &  &  &  &  &  &  & \textbf{0.01} & \textbf{0.00} \\
			2010-2015 &  &  &  &  &  &  &  &  &  & 0.92 \\
			\hline
		\end{tabular}
	}
	\caption{ASL values computed for the null hypothesis of time-homogeneity for the indicated sub-samples and 5-year sub-periods using the procedure described in the text. Authors' calculations using data from \citet{feenstra2015}.}
	\label{tab:testTimeHomogeneity5yearslowmediumhigh}
\end{table}

For the 5-year transition length, and both distance measures, we find considerably less evidence against the hypothesis of homogeneity for the low-income sub-sample than for the medium-income and high-income sub-samples. In Table \ref{tab:testTimeHomogeneity5yearslowmediumhigh} there are no rejections, even the 10\% significance level, of the homogeneity hypothesis for adjacent time periods for either distance measure for either the low-income or medium-income sub-samples, as inspection of the main diagonals of the relevant portions of the table reveals. For the medium-income countries, the homogeneity hypothesis is rejected at the 5\% level using the $L^1$ distance measure when 2005-2010 and 2010-2015 are compared with 1975-1980, 1985-1990, 1990-1995, and 1995-2000 and also when 2010-2015 is compared to those sub-periods and to 1970-1975. A similar pattern, albeit with fewer rejections, is observed using the $H$ distance measure. For the high-income countries there are rejections at the 5\% level using both distance measures of the hypothesis of homogenous dynamics when 1995-2000 is compared to 2000-2005 and to 2005-2010 and also when 2000-2005 and 2005-2010 are compared to 2010-2015 and to 2014-2019. That is, for the high-income countries in particular, the results are very similar to those for the sample as a whole reported in Table \ref{tab:testTimeHomogeneity5years} earlier suggesting a homogeneous process until the late 1990s after which there are apparent breaks in process in about 2000 and 2010.

\begin{table}[!htbp]
	\centering
	\begin{tabular}{rHrrrr}
		\hline
		\hline
		\textbf{$L^1$ Low income} & 1970-1980 & 1980-1990 & 1990-2000 & 2000-2010 & 2009-2019 \\
		\hline
		1970-1980 &  & 0.44 & 0.62 & 0.12 & 0.00 \\
		1980-1990 &  &  & 0.38 & 0.44 & 0.17 \\
		1990-2000 &  &  &  & 0.14 & 0.04 \\
		2000-2010 &  &  &  &  & 0.39 \\
	\end{tabular}
	\begin{tabular}{rHrrrr}
		\hline
		\textbf{$L^1$ Medium income} & 1970-1980 & 1980-1990 & 1990-2000 & 2000-2010 & 2009-2019 \\
		\hline
		1970-1980 &  & 0.43 & 0.01 & 0.01 & 0.00 \\
		1980-1990 &  &  & 0.34 & 0.09 & 0.01 \\
		1990-2000 &  &  &  & 0.13 & 0.03 \\
		2000-2010 &  &  &  &  & 0.61 \\
	\end{tabular}
	\begin{tabular}{rHrrrr}
		\hline
		\textbf{$L^1$ High income} & 1970-1980 & 1980-1990 & 1990-2000 & 2000-2010 & 2009-2019 \\
		\hline
		1970-1980 &  & 0.03 & 0.00 & 0.02 & 0.00 \\
		1980-1990 &  &  & 0.07 & 0.01 & 0.03 \\
		1990-2000 &  &  &  & 0.00 & 0.29 \\
		2000-2010 &  &  &  &  & 0.01 \\
	\end{tabular}
	%
	\begin{tabular}{rHrrrr}
		\hline
		\hline
		\textbf{$H$ Low income} & 1970-1980 & 1980-1990 & 1990-2000 & 2000-2010 & 2009-2019 \\
		\hline
		1970-1980 &  & 0.42 & 0.55 & 0.26 & 0.02 \\
		1980-1990 &  &  & 0.41 & 0.66 & 0.42 \\
		1990-2000 &  &  &  & 0.25 & 0.08 \\
		2000-2010 &  &  &  &  & 0.39 \\
	\end{tabular}
	\begin{tabular}{rHrrrr}
		\hline
		\textbf{$H$ Medium income} & 1970-1980 & 1980-1990 & 1990-2000 & 2000-2010 & 2009-2019 \\
		\hline
		1970-1980 &  & 0.40 & 0.02 & 0.02 & 0.00 \\
		1980-1990 &  &  & 0.21 & 0.09 & 0.01 \\
		1990-2000 &  &  &  & 0.18 & 0.04 \\
		2000-2010 &  &  &  &  & 0.52 \\
	\end{tabular}
	\begin{tabular}{rHrrrr}
		\hline
		\textbf{$H$ High income} & 1970-1980 & 1980-1990 & 1990-2000 & 2000-2010 & 2009-2019 \\
		\hline
		1970-1980 &  & 0.04 & 0.00 & 0.01 & 0.00 \\
		1980-1990 &  &  & 0.06 & 0.05 & 0.07 \\
		1990-2000 &  &  &  & 0.00 & 0.33 \\
		2000-2010 &  &  &  &  & 0.01 \\
		\hline
		\hline
	\end{tabular}
	\caption{ASL values computed for the null hypothesis of time-homogeneity for the indicated sub-samples and 10-year sub-periods using the procedure described in the text. Authors' calculations using data from \citet{feenstra2015}.}
	\label{tab:testTimeHomogeneity10yearslowmediumhigh}
\end{table}

In Table \ref{tab:testTimeHomogeneity10yearslowmediumhigh} we see similar results for the 10-year transition length, again for both distance measures. Inspection of the main diagonals of the relevant portions of the table reveals there are no rejections, even the 10\% significance level, of the homogeneity hypothesis for adjacent time periods for both distance measures in either the low-income or medium-income sub-samples. For the high-income countries, there are, however, rejections of homogeneity at the the 5\% level for three of the four comparisons of adjacent time periods and at the the 10\% level for the other one. For the low-income sub-sample, only the comparisons of 2009-2019 with 1970-1980 and with 1990-2000 yield rejections, in both cases at the 5\% level using the $L^1$ distance measure, and at the 5\% and 10\% levels respectively using the $H$ distance measure. For the medium-income countries, homogeneity is rejected at the 5\% level using both distance measures, when 1970-1980 is compared with 1990-2000, 2000-2010, and 2009-2019 and also when the latter period is compared with 1980-1990 and 1990-2000. Using the $L^1$ distance measure, for the high-income sub-sample, only two of the 10 pair-wise time-period comparisons don't yield rejections at the 10\% level, and only three don't yield rejections at the 10\% level using the $H$ distance measure. That is, for the medium-income and high-income countries, the results are very similar to those found for the sample as a whole reported in Table \ref{tab:testTimeHomogeneity10YearSubperiods} which suggesting a homogeneous process until the mid 1990s after which there is little evidence of homogeneity.

Keeping in mind the caveat due to the small samples in mind, these results suggest that the inhomogeneity in the process governing the evolution of the cross-country distribution of per capita GDP beginning in the mid-1990s documented in Section \ref{sec:timeHomogeneity} is driven primarily by the behaviour of the high-income and, perhaps to a lesser extent, medium-income countries. There is almost no evidence of such an inhomogeneity in the evolution of the low-income countries. This conclusion is entirely consistent with the behaviour of the cross-country distribution of per capita GDP evident in Figure \ref{fig:normalizedQuintileBoundaries} and Figure \ref{fig:estimatedCrossSectionalDistributionsDifferentYears}. It is also relevant to the claims of short-term $\beta$-convergence in the 2000s reported by \citet{patel2021new} and \citet{kremer2022converging} as it suggests that the evidence of "catching up" presented in those papers is driven by changes in the evolution of per capita GDP in high- and medium-income countries rather than by changes in the fortunes of low-income countries. The evolution of per capita GDP in the low-income countries in our sample seems to obey the same process for the entire 1970-2019 sample period and income levels in those countries exhibit no improvement relative to the sample average.

\section{Summary and conclusions}

Studies of the convergence hypothesis employing the distribution dynamics approach pioneered by \citet{quah1993empirical,quah1996empirics,quah1996twin,quah1996convergence, quah1997empirics,quah1999} have provided evidence of considerable persistence and a strong long-run tendency to multimodality in the cross-country distribution of per capita output consistent with the club convergence hypothesis. These studies typically assume that a time-homogeneous, first-order process describes the evolution of the cross-country distribution of per capita output. Using a panel of 102 countries from PWT 10.0, we have examined the veracity of that assumption for the 1970-2019 period. We first examined the homogeneity hypothesis by comparing transition kernels, the mapping from the distribution at a point in time to a future point following a transition, estimated for different sub-intervals of our data period. Homogeneity implies the constancy of the transition kernel across sub-periods. As the true transition length is unknown, we use transition lengths of 5, 10, 15, 20, and 25 years. We use bootstrap methods to assess the statistical significance of the the distance between the estimated transition kernels which we measure using four different metrics, as our Monte Carlo work suggests that the size and power properties of our tests depend on the choice of metric. We find that, for all the transition lengths that we use, we cannot reject the hypothesis of an homogeneous process for the 1970-1995 period. We also find evidence of breaks in the process in the late 1990s and in about 2010 suggesting that the process was homogeneous in each of the 2000-2010 and 2009-2019 periods.

For the three periods in which there is evidence of homogeneity, we exploit the relationship between the transition kernel for a single transition and that for two consecutive transitions implied by the Chapman-Kolmogorov equations for a first-order process, to test the hypothesis that the process is first-order. Again, this test requires the computation of the distance between distributions which we do using the same four different metrics as for the homogeneity tests. Again we assess the statistical significance of the observed deviations from the distance implied by the null hypothesis using bootstrap methods. We find evidence that the process was time-homogeneous and first-order in the 1970-1995 period for the 10-year transition length but not for the 5-year transition length. The 25-year sample span restricts our ability to test to these two transition lengths. This result suggests that, in terms of model specification at least, \cite{quah1996twin,quah1997empirics} was correct in using 15-year transition length for the analysis of the 1960-1985 period. For the 2000-2010 and 2010-2019 periods, we find evidence that the process was time-homogeneous and first-order for the 5-year transition length, the only one that we can test given the time span of the sub-samples but concerns about the power of the test due to the small sample sizes imply caution is warranted in drawing conclusions here.

We next investigate the transition dynamics within each of these periods by computing the ergodic distributions and their bootstrap confidence intervals implied by the estimated transition kernels for the 5-year and 10-year transition lengths. For the 1970-1995 period the two estimated ergodic distributions are very similar. Both are bimodal with both modes, and the antimode between them, occurring at very close to the same value of relative GDP per worker. These ergodic distributions are consistent with the existence of convergence clubs. For both of the transition lengths considered, the estimated ergodic densities for the 2000-2010 sub-period show clear unimodality, suggestive of a single convergence club and thus absolute convergence, consistent with the short-term $\beta$-convergence found by \citet{patel2021new} and \citet{kremer2022converging} for the same period. The differences in the (bimodal) estimated cross-country distributions of per capita GDP in 2000 and 2010 seem to reflect this tendency to convergence in that the mode  associated with the rich countries in 2010 lies slightly to the left of that in 2000. In the 2009-2019 sub-period, signs of bimodality, suggestive of divergence, re-emerge as the estimated ergodic distribution for the 5-year transition length is again bimodal and similar to those estimated for the 1970-1995 sub-period. The estimated cross-country distributions of per capita GDP in 2009 and 2019 are almost coincident suggesting that the tendency to convergence observed in the 2000s may have stopped.

Apart from underscoring the importance of testing the assumptions of time-homogeneity and first-orderedness, the finding of a shift in the transition dynamics from a period in which they appear to satisfy these assumptions and imply divergence, to a period in which they seem to imply convergence, and then to a period in which they seem to again imply divergence invites explanation. The large macroeconomic shocks of the late 1990s provide some clues. The 1997 Asian financial crisis which slowed growth in several SE Asian counties (albeit briefly), the rapid productivity growth in the US in 1995-2000 , the dot-com bubble of the late 1990s, and the housing bubble in the US and elsewhere in the early 2000s, all of which could have served to disrupt the dynamics of the cross-country distribution of per capita GDP and increase the gap between the already prosperous countries and the others, are candidate culprits for the end of the period of an apparently homogeneous process beginning in 1970 (or earlier) homogeneity that we find in the late 1990s. The subsequent collapse of the housing bubble in 2006, and the relatively deep and prolonged (by post-war standards) recession that it brought, served to slow the growth of the more developed countries in the 2000-2010 period allowing the developing countries to appear to "catch up" so that that the dynamics estimated for that period suggest convergence.\footnote{\citet{imam2025} attribute the inhomogeneity that they find in cross-country income dynamics in the mid-1990s to changes in the the number and severity of economic crises. Our tentative explanation is a little different as we view the large macroeconomic shocks and their differential effects as possible causes of both the inhomogeneities and the apparent associated changes in the convergence properties of the cross-country income dynamics that we have documented.}

\clearpage

\bibliographystyle{chicago}

\bibliography{references.bib}

\clearpage

\appendix

{\LARGE \textbf{Appendices} }

\renewcommand{\thefigure}{A\arabic{figure}}
\renewcommand{\thetable}{A\arabic{table}}
\setcounter{figure}{0}
\setcounter{table}{0}

\section{Monte Carlo experiments \label{app:MonteCarloExperiments}}

	In this Appendix we report the results of a few Monte Carlo experiments designed to explore the properties of our proposed testing procedures, in particular the effect of the different measures of divergence between distributions, on the empirical size and power of our tests. The estimation and bootstrap procedures that we simulate all follow the descriptions described in the text with the exception that we use the standard, rather than the adaptive, kernel density estimator in order to reduce the computational burden. We consider five alternative sample size $n$ of  50, 100, 200, 500 and 1000, with 2 transitions for each unit of observation, which implies samples of 100, 200, 400, 1000 and 2000 transitions overall. In each Monte Carlo run, the initial distribution of pseudo observations is drawn from a Gaussian distribution with zero mean and standard deviation equal to 0.15, i.e $y_{i0} \sim \mathcal{N}\left(0,0.15\right) $. The number of Monte Carlo runs for each setting is 1000, as is the number of bootstrap replications conducted for each Monte Carlo run. Based on these experiments, we conclude that while sample size is an important determinant of test properties, the measures of divergence $L^1$ and $H$ show the best empirical size and power performance for both the tests of homogeneity and of the order of the process.

	\subsection{The empirical size and power of the time-homogeneity test \label{sec:empiricalPowerTimeHomogeneity}}

	We evaluate the empirical size and power of our procedure for testing the hypothesis of time homogeneity using the following first-order auto-regressive process:
	\begin{equation}
		y_{it} = \begin{cases}
			\rho y_{it-1} + \epsilon_t \hspace{1.1cm } \text{ for } t=1,\cdots,102; \text{ and } i=1,\cdots,n,\\
			\left(\rho + \theta \right) y_{it-1} + \epsilon_{it}  \text{ for } t=103, \text{ and } i=1,\cdots,n,
		\end{cases}
		\label{eq:timeHomogeneityrMarkovProcess}
	\end{equation}
	with $\epsilon_{it} \sim \mathcal{N}\left(0,0.15\right)$, and $\rho \in (0.05,0.2,0.5,0.75)$ and $\theta \in (0, 0.05,0.10,0.15,0.20)$ (all the combinations satisfy the condition for stationarity $\rho + \theta <1$).
	We discard the first 100 periods so that the process is in its long-run stochastic steady state. Then, the first transition for each observational unit is governed by the parameter $\rho$, while the second transition by the parameter $\rho+\theta$. Table \ref{tab:empiricalPowerTimeHomogeneity} reports the percentage of rejections of the null hypotheses when pseudo observations are generated by Equation (\ref{eq:timeHomogeneityrMarkovProcess}) with a rejection occurring if the $ASL\,<0.05$ so that the nominal test size is 5\%.
For each sample size, the second column of Table \ref{tab:empiricalPowerTimeHomogeneity} reports the result for $\theta=0$, i.e. the empirical size of the test. We observe values close to 0.05 even for the sample size of 100, especially for $L^1$ and $H$. The empirical power is above 0.5 for a sample size of 100 when the measure of divergence is $L^1$ or $H$ but only for the extreme case of $\theta = 0.5$. This rises to about 0.8 for a sample size of 200. For a moderate time inhomogeneity, i.e. $\theta = 0.25$, the empirical power is a good deal lower, falling to about 0.25. It is only for a sample of 1000 that it rises to about 0.75.  Overall, the measures of divergence $L^1$ and $H$ show the best empirical size and power performance, but the sample size appears to be a crucial determinant of the empirical power in particular.

	\begin{longtable}{llllll}
			\hline
			\hline
			$\rho$/$\theta$ & 0 & 0.05 & 0.15 & 0.25 & 0.5 \\
			\\
			\hline
			\multicolumn{6}{c}{\textbf{Sample size 50}} \\
			\multicolumn{6}{l}{$\mathbf{L^1}$}  \\
			0.05 & 0.032 & 0.037 & 0.06 & 0.09 & 0.304 \\
			0.2 & 0.034 & 0.021 & 0.05 & 0.089 & 0.34 \\
			\multicolumn{6}{l}{$\mathbf{L^2}$}  \\
			0.05 & 0.027 & 0.026 & 0.038 & 0.068 & 0.229 \\
			0.2 & 0.02 & 0.012 & 0.033 & 0.061 & 0.25 \\
			\multicolumn{6}{l}{$\mathbf{L^\infty}$}  \\
			0.05 & 0.018 & 0.011 & 0.009 & 0.012 & 0.031 \\
			0.2 & 0.008 & 0.01 & 0.013 & 0.014 & 0.047 \\
			\multicolumn{6}{l}{$\mathbf{H}$}  \\
			0.05 & 0.042 & 0.037 & 0.049 & 0.099 & 0.352 \\
			0.2 & 0.032 & 0.028 & 0.059 & 0.11 & 0.383 \\
			\\
			\hline
			\multicolumn{6}{c}{\textbf{Sample size 100}} \\
			\multicolumn{6}{l}{$\mathbf{L^1}$}  \\
			0.05 &  0.046 &0.038&  0.08& 0.155& 0.575 \\
			0.2   & 0.053 &0.047& 0.074& 0.141& 0.558 \\
			\multicolumn{6}{l}{$\mathbf{L^2}$}  \\
			0.05 & 0.028& 0.016& 0.054& 0.098& 0.405 \\
			0.2  & 0.035 &0.031& 0.047& 0.103& 0.377 \\
			\multicolumn{6}{l}{$\mathbf{L^\infty}$}  \\
			0.05 & 0.013& 0.012& 0.013& 0.021 &0.076 \\
			0.2  & 0.008& 0.012& 0.016& 0.039 &0.093 \\
			\multicolumn{6}{l}{$\mathbf{H}$}  \\
			0.05 &0.045 &0.027& 0.082& 0.152& 0.561 \\
			0.2  & 0.046 & 0.045  & 0.08& 0.157& 0.559\\
			\\
			\hline
			\multicolumn{6}{c}{\textbf{Sample size 200}} \\
			\multicolumn{6}{l}{$\mathbf{L^1}$}  \\
			0.05 &  0.047 & 0.059 & 0.12 & 0.253 & 0.854\\
			0.2  &  0.049 & 0.054 &  0.097 & 0.258 & 0.837\\
			\multicolumn{6}{l}{$\mathbf{L^2}$}  \\
			0.05 & 0.035 & 0.037 & 0.059 & 0.149 & 0.586\\
			0.2  & 0.031 & 0.03 & 0.056 & 0.146 & 0.553\\
			\multicolumn{6}{l}{$\mathbf{L^\infty}$}  \\
			0.05 & 0.015 & 0.024 & 0.022 & 0.034 & 0.105\\
			0.2  & 0.028 & 0.013 &0.033 & 0.043 & 0.132\\
			\multicolumn{6}{l}{$\mathbf{H}$}  \\
			0.05 & 0.036 & 0.046 & 0.103 & 0.226 & 0.793\\
			0.2  & 0.041 & 0.048 &0.087 & 0.232  & 0.77\\
			\\
			\hline
			\multicolumn{6}{c}{\textbf{Sample size 1000}} \\
			\multicolumn{6}{l}{$\mathbf{L^1}$}  \\
			0.05 & 0.048 & 0.08 & 0.274 & 0.765 & 1 \\
			0.2 & 0.058 & 0.08 & 0.279 & 0.74 & 1 \\
			\multicolumn{6}{l}{$\mathbf{L^2}$}  \\
			0.05 & 0.045 & 0.052 & 0.123 & 0.391 & 0.989 \\
			0.2 & 0.041 & 0.051 & 0.099 & 0.301 & 0.972 \\
			\multicolumn{6}{l}{$\mathbf{L^\infty}$}  \\
			0.05 & 0.032 & 0.03 & 0.035 & 0.062 & 0.164 \\
			X0.2 & 0.029 & 0.025 & 0.039 & 0.055 & 0.129 \\
			\multicolumn{6}{l}{$\mathbf{H}$}  \\
			0.05 & 0.059 & 0.07 & 0.209 & 0.612 & 0.998 \\
			0.2 & 0.061 & 0.065 & 0.175 & 0.519 & 0.999 \\
			\hline
			\hline
			\caption{The percentage of rejections of the null hypothesis of a time-homogeneous process as given in (\ref{test:homogeneity}) for a test with a nominal size of 5\% for different $\rho$ and $\theta$, measures of divergence given in Equations (\ref{eq:LpNorm}) and (\ref{eq:HillengerDistance}), and sample size. The densities are estimated by a kernel density estimator using a Gaussian kernel and the optimal normal bandwidth. The latter is constant for each bootstrap replication conducted for a Monte Carlo iteration but is re-estimated for each Monte Carlo iteration using the data drawn for that iteration. The range of estimation is $\left[-3\times \left[0.15_{\epsilon}/(1-(\rho+\theta)^2) \right]^{1/2},3\times \left[0.15_{\epsilon}/(1-(\rho+\theta)^2) \right]^{1/2} \right]$ in a regular grid of $40 \times 40$ evaluation points.}
			\label{tab:empiricalPowerTimeHomogeneity}
		\end{longtable}

	\subsection{The empirical size and power of the test on the order of process \label{sec:empiricalPowerFirstOrder}}

	We evaluate the empirical size and power of our procedure for testing the hypothesis of a first-order Markov process using the following second-order auto-regressive process:
	\begin{equation}
		y_{it} = \rho_1 y_{it-1} + \rho_2 y_{it-2} + \epsilon_{it}  \text{ for } t=1,\cdots,103, \text{ and } i=1,\cdots,n,
		\label{eq:firstVsSecondOrderMarkovProcess}
	\end{equation}
	with  $\epsilon_{it} \sim \mathcal{N}\left(0,0.15\right)$, $\rho_1 \in \{ 0.2,0.5,0.75\}$ and $\rho_2 \in \{0, -0.10,-0.25 \}$. As $|\rho_1 + \rho_2|<1$ the generated series is always stationary. Again, we discard the first 100 periods so that the process is in its long-run stochastic steady state.
	Table \ref{tab:empiricalPowerOrderProcessAlternative} reports the percentage of rejections of the null hypotheses when pseudo observations are generated by Equation (\ref{eq:firstVsSecondOrderMarkovProcess}) with a rejection occurring if the $ASL\,<0.05$ so that the nominal test size is 5\%.
	For each sample size, the second column of Table \ref{tab:empiricalPowerOrderProcessAlternative} reports the result for $\rho_2=0$, i.e. the empirical size of the test. We observe values close to 0.05 even for the sample size of 100 for $\rho_1 =0.2$ and for $\rho_1 =0.5$, with $L^1$ and $H$ performing a bit better than the other two divergence measures. However, for $\rho_1 =0.75$, both of these divergence measures display an empirical size well below the nominal size of 0.05 while $L^2$ and $L^\infty$ perform better when the persistence in the process increases. As far is the empirical power is concerned, $L^1$,  $L^2$, and $H$ generally perform be than $L^\infty$, although none of the divergence measures produces high power for small deviations from the null hypothesis (i.e $\rho_2=-0.1$), even for a sample size of 1000. Overall, the measures of divergence $L^1$ and $H$ show the best empirical size and power performance, but the sample size appears to be a crucial determinant of the empirical power in particular.

	\begin{longtable}{llll}
			\hline
			\hline
			$\rho_1/\rho_2$   & 0 & -0.1 & -0.25 \\
			\\
			\hline
			\multicolumn{4}{c}{\textbf{Sample size 50}} \\
			\multicolumn{4}{l}{$\mathbf{L^1}$}  \\
			0.2 & 0.05 & 0.078 & 0.217 \\
			0.5 & 0.055 & 0.078 & 0.217 \\
			0.75 & 0.003 & 0.026 & 0.178 \\
			\multicolumn{4}{l}{$\mathbf{L^2}$}  \\
			0.2 & 0.045 & 0.07 & 0.177 \\
			0.5 & 0.069 & 0.087 & 0.204 \\
			0.75 & 0.01 & 0.052 & 0.247 \\
			\multicolumn{4}{l}{$\mathbf{L^\infty}$}  \\
			0.2 & 0.038 & 0.041 & 0.054 \\
			0.5 & 0.06 & 0.072 & 0.123 \\
			0.75 & 0.036 & 0.078 & 0.145 \\
			\multicolumn{4}{l}{$\mathbf{H}$}  \\
			0.2 & 0.032 & 0.062 & 0.2 \\
			0.5 & 0.0591 & 0.056 & 0.169 \\
			0.75 & 0.006 & 0.039 & 0.218 \\
			\\
			\hline
			\multicolumn{4}{c}{\textbf{Sample size 100}} \\
			\multicolumn{4}{l}{$\mathbf{L^1}$}  \\
			0.2 & 0.059 & 0.084 & 0.325 \\
			0.5 & 0.062 & 0.104 & 0.309 \\
			0.75 & 0.005 & 0.033 & 0.265 \\
			\multicolumn{4}{l}{$\mathbf{L^2}$}  \\
			0.2 & 0.054 & 0.08 & 0.24 \\
			0.5 & 0.078 & 0.127 & 0.281 \\
			0.75 & 0.014 & 0.056 & 0.358 \\
			\multicolumn{4}{l}{$\mathbf{L^\infty}$}  \\
			0.2 & 0.06 & 0.068 & 0.081 \\
			0.5 & 0.085 & 0.118 & 0.149 \\
			0.75 & 0.031 & 0.089 & 0.316 \\
			\multicolumn{4}{l}{$\mathbf{H}$}  \\
			0.2 & 0.045 & 0.065 & 0.275 \\
			0.5 & 0.062 & 0.074 & 0.221 \\
			0.75 & 0.007 & 0.041 & 0.298 \\
			\\
			\hline
			\multicolumn{4}{c}{\textbf{Sample size 200}} \\
			\multicolumn{4}{l}{$\mathbf{L^1}$}  \\
			0.2 & 0.057 & 0.098 & 0.483 \\
			0.5 & 0.058 & 0.117 & 0.358 \\
			0.75 & 0.016 & 0.046 & 0.329 \\
			\multicolumn{4}{l}{$\mathbf{L^2}$}  \\
			0.2 & 0.056 & 0.089 & 0.323 \\
			0.5 & 0.078 & 0.124 & 0.287 \\
			0.75 & 0.034 & 0.099 & 0.433 \\
			\multicolumn{4}{l}{$\mathbf{L^\infty}$}  \\
			0.2 & 0.055 & 0.066 & 0.094 \\
			0.5 & 0.088 & 0.109 & 0.151 \\
			0.75 & 0.062 & 0.126 & 0.33 \\
			\multicolumn{4}{l}{$\mathbf{H}$}  \\
			0.2 & 0.047 & 0.077 & 0.408 \\
			0.5 & 0.068 & 0.085 & 0.249 \\
			0.75 & 0.02 & 0.054 & 0.35 \\
			\\
			\hline
			\multicolumn{4}{c}{\textbf{Sample size 1000}} \\
			\multicolumn{4}{l}{$\mathbf{L^1}$}  \\
			0.2 & 0.057 & 0.179 & 0.963 \\
			0.5 & 0.058 & 0.156 & 0.589 \\
			0.75 & 0.019 & 0.062 & 0.429 \\
			\multicolumn{4}{l}{$\mathbf{L^2}$}  \\
			0.2 & 0.044 & 0.097 & 0.662 \\
			0.5 & 0.083 & 0.147 & 0.346 \\
			0.75 & 0.045 & 0.116 & 0.499 \\
			\multicolumn{4}{l}{$\mathbf{L^\infty}$}  \\
			0.2 & 0.047 & 0.053 & 0.066 \\
			0.5 & 0.12 & 0.123 & 0.18 \\
			0.75 & 0.063 & 0.115 & 0.346 \\
			\multicolumn{4}{l}{$\mathbf{H}$}  \\
			0.2 & 0.04 & 0.109 & 0.866 \\
			0.5 & 0.073 & 0.107 & 0.309 \\
			0.75 & 0.015 & 0.061 & 0.403 \\
			\hline
			\hline
		\caption{The percentage of rejections of the null hypothesis of a first-order Markov process as given in (\ref{test:ChapmanKolmogorov}) for a test with a nominal size of 5\% for different $\rho_1$ and $\rho_2$, measures of divergence given in Equations (\ref{eq:LpNorm}) and (\ref{eq:HillengerDistance}), and sample size. The densities are estimated by a kernel density estimator using a Gaussian kernel and the optimal normal bandwidth. The latter is constant for each bootstrap replication conducted for a Monte Carlo iteration but is re-estimated for each Monte Carlo iteration using the data drawn for that iteration. The range of estimation is $\left[-3\times \left[0.15_{\epsilon}/(1-(\rho_1 +\rho_2)^2) \right]^{1/2},3\times \left[0.15_{\epsilon}/(1-(\rho_1 +\rho_2)^2) \right]^{1/2} \right]$ in a regular grid of $40 \times 40$ evaluation points.}
		\label{tab:empiricalPowerOrderProcessAlternative}
	\end{longtable}

\clearpage

%
%

\renewcommand{\thefigure}{B\arabic{figure}}
\renewcommand{\thetable}{B\arabic{table}}
\setcounter{figure}{0}
\setcounter{table}{0}

\section{Time-homogeneity tests using all divergence measures}
\label{app:testingTimeHomogeneity}

\subsection{5-year transitions}
\begin{table}[!htbp]
	\centering
	\hspace{-1cm}
	\scriptsize{
		\begin{tabular}{rHrrrrrrrrr}
			\hline
			\hline
			\textbf{Test based on $L^1$} & 1970-1975 & 1975-1980 & 1980-1985 & 1985-1990 & 1990-1995 & 1995-2000 & 2000-2005 & 2005-2010 & 2010-2015 & 2014-2019 \\
			\hline
			1970-1975 &  & 0.92 & 0.40 & 0.44 & 0.96 & \textbf{0.00} & 0.17 & \textbf{0.01} & \textbf{0.00} & \textbf{0.00} \\
			1975-1980 &  &  & 0.18 & 0.64 & 0.85 & \textbf{0.00 }& 0.19 & \textbf{0.00} & \textbf{0.00} & \textbf{0.01} \\
			1980-1985 &  &  &  & 0.06 & 0.34 & \textbf{0.00} & 0.65 & 0.10 & \textbf{0.01} & \textbf{0.03} \\
			1985-1990 &  &  &  &  & 0.40 & \textbf{0.00} & \textbf{0.01 }& \textbf{0.00} & \textbf{0.00 }& \textbf{0.00} \\
			1990-1995 &  &  &  &  &  & \textbf{0.02} & 0.21 & \textbf{0.00} & \textbf{0.00 }&\textbf{ 0.01} \\
			1995-2000 &  &  &  &  &  &  & \textbf{0.00} & \textbf{0.00} & \textbf{0.00} & 0.33 \\
			2000-2005 &  &  &  &  &  &  &  & 0.10 & \textbf{0.02} & \textbf{0.04} \\
			2005-2010 &  &  &  &  &  &  &  &  & \textbf{0.01} & \textbf{0.00 }\\
			2010-2015 &  &  &  &  &  &  &  &  &  & 0.19 \\
			\hline
			\\
			\textbf{Test based on $L^2$} & 1970-1975 & 1975-1980 & 1980-1985 & 1985-1990 & 1990-1995 & 1995-2000 & 2000-2005 & 2005-2010 & 2010-2015 & 2014-2019 \\
			\hline
			1970-1975 &  & 0.92 & 0.26 & 0.54 & 0.92 & 0.01 & 0.13 & 0.01 & 0.00 & 0.01 \\
			1975-1980 &  &  & 0.16 & 0.69 & 0.78 & 0.01 & 0.11 & 0.00 & 0.00 & 0.00 \\
			1980-1985 &  &  &  & 0.06 & 0.12 & 0.02 & 0.74 & 0.25 & 0.02 & 0.05 \\
			1985-1990 &  &  &  &  & 0.40 & 0.00 & 0.01 & 0.00 & 0.00 & 0.00 \\
			1990-1995 &  &  &  &  &  & 0.06 & 0.07 & 0.00 & 0.00 & 0.01 \\
			1995-2000 &  &  &  &  &  &  & 0.00 & 0.00 & 0.00 & 0.52 \\
			2000-2005 &  &  &  &  &  &  &  & 0.23 & 0.02 & 0.03 \\
			2005-2010 &  &  &  &  &  &  &  &  & 0.02 & 0.00 \\
			2010-2015 &  &  &  &  &  &  &  &  &  & 0.15 \\
		\hline
		\\
		\textbf{Test based on $L^\infty$}& 1970-1975 & 1975-1980 & 1980-1985 & 1985-1990 & 1990-1995 & 1995-2000 & 2000-2005 & 2005-2010 & 2010-2015 & 2014-2019 \\
		\hline
		1970-1975 &  & 0.83 & 0.19 & 0.76 & 0.88 & 0.05 & 0.20 & 0.11 & 0.01 & 0.04 \\
		1975-1980 &  &  & 0.19 & 0.96 & 0.48 & 0.01 & 0.10 & 0.06 & 0.00 & 0.00 \\
		1980-1985 &  &  &  & 0.16 & 0.08 & 0.26 & 0.79 & 0.64 & 0.10 & 0.28 \\
		1985-1990 &  &  &  &  & 0.40 & 0.01 & 0.09 & 0.05 & 0.01 & 0.01 \\
		1990-1995 &  &  &  &  &  & 0.05 & 0.08 & 0.01 & 0.00 & 0.04 \\
		1995-2000 &  &  &  &  &  &  & 0.14 & 0.06 & 0.02 & 0.52 \\
		2000-2005 &  &  &  &  &  &  &  & 0.30 & 0.07 & 0.30 \\
		2005-2010 &  &  &  &  &  &  &  &  & 0.17 & 0.10 \\
		2010-2015 &  &  &  &  &  &  &  &  &  & 0.25 \\
			\hline
			\\
			\textbf{Test based on $H$} & 1970-1975 & 1975-1980 & 1980-1985 & 1985-1990 & 1990-1995 & 1995-2000 & 2000-2005 & 2005-2010 & 2010-2015 & 2014-2019 \\
			\hline
			1970-1975 &  & 0.86 & 0.35 & 0.29 & 0.99 & \textbf{0.00} & 0.46 &\textbf{ 0.02} & \textbf{0.01} & \textbf{0.02} \\
			1975-1980 &  &  & 0.14 & 0.71 & 0.89 & \textbf{0.00} & 0.31 & \textbf{0.00 }& \textbf{0.00 }& \textbf{0.04} \\
			1980-1985 &  &  &  & 0.09 & 0.24 & \textbf{0.01} & 0.31 & 0.11 & 0.10 & 0.12 \\
			1985-1990 &  &  &  &  & 0.45 & \textbf{0.01 }& \textbf{0.03} & \textbf{0.00 }& \textbf{0.01 }& \textbf{0.03} \\
			1990-1995 &  &  &  &  &  & \textbf{0.01} & 0.38 & \textbf{0.00} & \textbf{0.01 }& 0.16 \\
			1995-2000 &  &  &  &  &  &  & \textbf{0.00 }& \textbf{0.00 }& \textbf{0.00} & 0.22 \\
			2000-2005 &  &  &  &  &  &  &  & 0.06 & 0.07 & 0.07 \\
			2005-2010 &  &  &  &  &  &  &  &  & \textbf{0.01} & \textbf{0.00} \\
			2010-2015 &  &  &  &  &  &  &  &  &  & 0.51 \\
			\hline
			\\
			Number of obs & 102 & 102 & 102 & 102 & 102 & 102 & 102 & 102 & 102  & 102 \\
			\hline
			\hline
		\end{tabular}
	}
	\caption{ASL values computed for the null hypothesis of time-homogeneity for the indicated 5-year sub-periods using the procedure described in the text. Authors' calculations using data from \citet{feenstra2015}.}
	\label{tab:testTimeHomogeneity5years_app}
\end{table}

\clearpage

\subsection{10-year transitions}

\begin{table}[!htbp]
	\centering
	\begin{tabular}{rHrrrr}
		\hline
		\hline
		\textbf{Test based on $L^1$}  & 1970-1980 & 1980-1990 & 1990-2000 & 2000-2010 & 2009-2019 \\
		\hline
		1970-1980 &  & 0.38 & \textbf{0.00 }& \textbf{0.00 }& \textbf{0.00} \\
		1980-1990 &  &  & 0.06 & \textbf{0.00} & \textbf{0.01} \\
		1990-2000 &  &  &  & \textbf{0.00 }& \textbf{0.01} \\
		2000-2010 &  &  &  &  & \textbf{0.03}\\
		\hline
		\\
		\textbf{Test based on $L^2$}   & 1970-1980 & 1980-1990 & 1990-2000 & 2000-2010 & 2009-2019 \\
		\hline
		1970-1980 &  & 0.47 & 0.01 & 0.00 & 0.00 \\
		1980-1990 &  &  & 0.07 & 0.00 & 0.0 \\
		1990-2000 &  &  &  & 0.00 & 0.00 \\
		2000-2010 &  &  &  &  & 0.02 \\
		\hline
		\\
		\textbf{Test based on $L^\infty$}  & 1970-1980 & 1980-1990 & 1990-2000 & 2000-2010 & 2009-2019 \\
		\hline
		1970-1980 &  & 0.42 & 0.02 & 0.09 & 0.00 \\
		1980-1990 &  &  & 0.08 & 0.20 & 0.04 \\
		1990-2000 &  &  &  & 0.01 & 0.01 \\
		2000-2010 &  &  &  &  & 0.28 \\
		\hline
		\\
		\textbf{Test based on $H$}  & 1970-1980 & 1980-1990 & 1990-2000 & 2000-2010 & 2009-2019 \\
		\hline
		1970-1980 &  & 0.43 & \textbf{0.01} & \textbf{0.01} & \textbf{0.01} \\
		1980-1990 &  &  & 0.07 & \textbf{0.01} & \textbf{0.04} \\
		1990-2000 &  &  &  & \textbf{0.00 }& \textbf{0.03} \\
		2000-2010 &  &  &  &  & \textbf{0.04} \\
		\hline
		\\
		Number of obs & 102 & 102 & 102 & 102 & 102 \\
		\hline
		\hline
	\end{tabular}
	\caption{ASL values computed for the null hypothesis of time-homogeneity for the indicated 10-year sub-periods using the procedure described in the text. Authors' calculations using data from \citet{feenstra2015}.}
	\label{tab:testTimeHomogeneity10YearSubperiods_app}
\end{table}

\clearpage

\subsection{15-year transitions}

\begin{table}[!htbp]
	\begin{tabular}{rHrr}
		\hline
		\hline
		\textbf{Test based on $L^1$}  & 1970-1985 & 1985-2000 & 2000-2015 \\
		\hline
		1970-1985 &  & 0.15 & \textbf{0.01} \\
		1985-2000 &  &  & \textbf{0.0}0 \\
		\hline
		\\
		\textbf{Test based on $L^2$} & 1970-1985 & 1985-2000 & 2000-2015 \\
		\hline
		1970-1985 &  & 0.65 & 0.05 \\
		1985-2000 &  &  & 0.00 \\
		\hline
		\\
		\textbf{Test based on $L^\infty$} & 1970-1985 & 1985-2000 & 2000-2015 \\
		\hline
		1970-1985 &  & 0.65 & 0.05 \\
		1985-2000 &  &  & 0.00 \\
		\hline
		\\
		\textbf{Test based on $H$} & 1970-1985 & 1985-2000 & 2000-2015 \\
		\hline
		1970-1985 &  & 0.23 & \textbf{0.02} \\
		1985-2000 &  &  & \textbf{0.00} \\
		\hline
		\\
		Number of obs & 102 & 102 & 102 \\
		\hline
		\hline
	\end{tabular}
	\begin{tabular}{rHrr}
		\hline
		\hline
		\textbf{Test based on $L^1$}  & 1974-1989 & 1989-2004 & 2004-2019 \\
		\hline
		1974-1989 &  & 0.06 & \textbf{0.00} \\
		1989-2004 &  &  & \textbf{0.00 }\\
		\hline
		\\
		\textbf{Test based on $L^2$}  & 1974-1989 & 1989-2004 & 2004-2019 \\
		\hline
		1974-1989 &  & 0.10 & {0.00} \\
		1989-2004 &  &  & 0.00 \\
		\hline
		\hline
		\\
		\textbf{Test based on $L^\infty$}   & 1974-1989 & 1989-2004 & 2004-2019 \\
		\hline
		1974-1989 &  & 0.08 & 0.07 \\
		1989-2004 &  &  & 0.04 \\
		\hline
		\\
		\textbf{Test based on $H$}   & 1974-1989 & 1989-2004 & 2004-2019 \\
		\hline
		1974-1989 &  & 0.07 & \textbf{0.01} \\
		1989-2004 &  &  & \textbf{0.03} \\
		\hline
		\\
		Number of obs & 102 & 102 & 102 \\
		\hline
		\hline
	\end{tabular}
	\caption{ASL values computed for the null hypothesis of time-homogeneity for the indicated 15-year sub-periods using the procedure described in the text. Authors' calculations using data from \citet{feenstra2015}.}
	\label{tab:testTimeHomogeneity15YearSubperiods_app}
\end{table}

\clearpage

\subsection{20-year transitions}

\begin{table}[!htbp]
	\centering
	\begin{tabular}{rHr}
		\hline
		\hline
		\textbf{Test based on $L^1$} & 1970-1990 & 1990-2010 \\
		\hline
		1970-1990 &  & \textbf{0.00} \\
		\hline
		\\
		\textbf{Test based on $L^2$}  & 1970-1990 & 1990-2010 \\
		\hline
		1970-1990 &  & 0.01 \\
		\hline
		\\
		\textbf{Test based on $L^\infty$} & 1970-1990 & 1990-2010 \\
		\hline
		1970-1990 &  & 0.29 \\
		\\
		\textbf{Test based on $H$}  & 1970-1990 & 1990-2010 \\
		\hline
		1970-1990 &  & \textbf{0.01} \\
		\hline
		\\
		Number of obs & 102 & 102\\
		\hline
		\hline
	\end{tabular}
	\begin{tabular}{rHr}
		\hline
		\hline
		\textbf{Test based on $L^1$}    & 1979-1999 & 1999-2019 \\
		\hline
		1979-1999 &  & \textbf{0.00} \\
		\hline
		\\
		\textbf{Test based on $L^2$}    & 1979-1999 & 1999-2019 \\
		\hline
		1979-1999 &  & 0.00 \\
		\hline
		\\
		\textbf{Test based on $L^\infty$}    & 1979-1999 & 1999-2019 \\
		\hline
		1979-1999 &  & 0.02 \\
		\\
		\textbf{Test based on $H$}    & 1979-1999 & 1999-2019 \\
		\hline
		1979-1999 &  &\textbf{ 0.00} \\
		\hline
		\\
		Number of obs & 102 & 102\\
		\hline
		\hline
	\end{tabular}
	\caption{ASL values computed for the null hypothesis of time-homogeneity for the indicated 20-year sub-periods using the procedure described in the text. Authors' calculations using data from \citet{feenstra2015}.}
	\label{tab:testTimeHomogeneity20YearSubperiods_app}
\end{table}

\clearpage

\subsection{25-year transitions}

\begin{table}[!htbp]
	\centering
	\begin{tabular}{rHr}
		\hline
		\hline
		\textbf{Test based on $L^1$}    & 1970-1995 & 1994-2019 \\
		\hline
		1970-1995 &  & \textbf{0.00 }\\
		\hline
		\\
		\textbf{Test based on $L^2$}    & 1970-1995 & 1994-2019 \\
		\hline
		1970-1995 &  & 0.00 \\
		\hline
		\\
		\textbf{Test based on $L^\infty$}    & 1970-1995 & 1994-2019 \\
		\hline
		1970-1995&  & 0.01 \\
		\hline
		\\
		\textbf{Test based on $H$}  & 1970-1995 & 1994-2019 \\
		\hline
		1970-1995 &  & \textbf{0.00} \\
		\hline
		\\
		Number of obs & 102 & 102\\
		\hline
		\hline
	\end{tabular}
	\caption{ASL values computed for the null hypothesis of time-homogeneity for the indicated 25-year sub-periods using the procedure described in the text. Authors' calculations using data from \citet{feenstra2015}.}
	\label{tab:testTimeHomogeneity24YearSubperiods_app}

\end{table}

\clearpage

\end{document}